\documentclass[12pt, authoryear]{elsarticle}              
\journal{Advances in Econometrics}            

\usepackage{amssymb,amsfonts,amsmath}
\usepackage{xcolor}
\usepackage{natbib}
\usepackage[top=1.25in, bottom=1.25in, left=1.0in, right=1.0 in]{geometry}   
\usepackage[onehalfspacing]{setspace}
\usepackage[bottom]{footmisc}
\usepackage{indentfirst}
\usepackage{endnotes}
\usepackage{verbatim}                
\usepackage{lastpage}
\usepackage{hyperref}                
\usepackage{epsfig,pgf,subfig}       
\usepackage{array}
\usepackage{rotating,longtable,booktabs, colortbl,siunitx}       
\usepackage{mdwlist,enumitem}
\usepackage{lscape,afterpage}                    
\usepackage{appendix}
\usepackage{breakcites}

\usepackage[standard]{ntheorem}
\usepackage[algo2e]{algorithm2e}

\usepackage[T1]{fontenc}
\usepackage{lmodern}

\allowdisplaybreaks[1]                     
\theoremstyle{break}
\theorembodyfont{\normalfont}
\newtheorem{algorithm}[algocf]{Algorithm}

\newcommand{\ack}[1]{%
  \nonumnote{\textit{Acknowledgements:\enspace}#1}}

\makeatletter    
\def\@author#1{\g@addto@macro\elsauthors{\normalsize%
    \def\baselinestretch{1}%
    \upshape\authorsep#1\unskip\textsuperscript{%
      \ifx\@fnmark\@empty\else\unskip\sep\@fnmark\let\sep=,\fi
      \ifx\@corref\@empty\else\unskip\sep\@corref\let\sep=,\fi
      }%
    \def\authorsep{\unskip,\space}%
    \global\let\@fnmark\@empty
    \global\let\@corref\@empty  
    \global\let\sep\@empty}%
    \@eadauthor={#1}
}

\makeatother     

\begin{document}

\begin{frontmatter}
\title{\textbf{Estimation and Applications of Quantile Regression for
Binary Longitudinal Data}}


\author{Mohammad Arshad Rahman\fnref{fn1}}\fntext[fn1]{Room 672, Faculty Building,
IIT Kanpur, Kanpur 208016, India; \emph{email: marshad@iitk.ac.in}.}

\address{Department of Economic Sciences, Indian Institute of Technology Kanpur}

\author{Angela Vossmeyer\fnref{fn2}}\fntext[fn2]{Robert Day School of Economics and
Finance, Claremont McKenna College, 500 E. Ninth St., Claremont, CA 91711;
\emph{email: angela.vossmeyer@cmc.edu}.}

\address{Robert Day School of Economics and
Finance, Claremont McKenna College}

\ack{The authors thank the anonymous referees, Ivan Jeliazkov, David
Brownstone, John Geweke, K.L. Krishna, Antonio Galvao, Michael Guggisberg,
and Editor Justin Tobias for their helpful comments. Discussions and
suggestions from the participants at the Winter School, Delhi School of
Economics (2017) and Advances in Econometrics Conference (2018) are
appreciated. A special thanks to Dale Poirier for sharing the Reverend's
insights and teaching us the controversy.}


\begin{abstract}
This paper develops a framework for quantile regression in binary
longitudinal data settings. A novel Markov chain Monte Carlo (MCMC) method
is designed to fit the model and its computational efficiency is
demonstrated in a simulation study. The proposed approach is flexible in
that it can account for common and individual-specific parameters, as well
as multivariate heterogeneity associated with several covariates. The
methodology is applied to study female labor force participation and home
ownership in the United States. The results offer new insights at the
various quantiles, which are of interest to policymakers and researchers
alike.

\end{abstract}

\end{frontmatter}

\section{Introduction}\label{sec:Intro}

The proliferation of panel data studies is well-documented and much of it has
been attributed to data availability and challenging methodology
\citep{Hsiao-2007}. While panel data has been attractive for understanding
behavior and dynamics, the modeling complexities involved in it have moved
attention away from its unique capacities. Modeling features such as a binary
outcome variable or a quantile analysis, which are relatively straightforward
to implement with cross-sectional data, are challenging and computationally
burdensome for panel data. However, these features are important as they
allow for the modeling of probabilities and lead to a richer view of how the
covariates influence the outcome variable. Motivated by these difficulties,
this paper adds to the methodological advancements for panel data by
developing quantile regression methods for binary longitudinal data and
designing a computationally efficient estimation algorithm. The approach is
applied to two empirical studies, female labor force participation and home
ownership.

The paper touches on three growing econometric literatures -- discrete panel
data, quantile regression for panel data, and quantile regression for
discrete data. In reference to the latter, quantile regression has been
implemented in binary data models \citep{Kordas-2006,Benoit-Poel-2012},
ordered data models \citep{Rahman-2016,Alhamzawi-Ali-2018}, count data models
\citep{MachadoSilva2005,HardingLamarche2015}, and censored data models
\citep{Portnoy2003,HardingLamarche2012}. For limited dependent variables, the
concern is modeling the latent utility differential in the quantile
framework, since the response variable takes limited values and does not
yield continuous quantiles. Our paper follows the work in this literature by
using the latent utility setting and interpreting the utility as a
``propensity'' or ``willingness'' that underlie the latent scale, thus
increasing our understanding of the impact of the covariates on the binary
outcomes.

\begin{sloppypar}
The literature on quantile regression in panel data settings
includes (but is not limited to) \citet{Koenker-2004},
\citet{Geraci-Bottai-2007}, \citet{Liu-Bottai-2009}, \citet{Galvao2010},
\citet{GalvaoKato2016}, \citet{Lamarche2010}, \citet{HardingLamarche2009} and
\citet{HardingLamarche2017}. The latter of these papers discusses the issues
associated with solely focusing on fixed effects estimators and highlights
the usefulness of allowing for a flexible specification of individual
heterogeneity associated with covariates, also of interest in the present
paper. In a recent Bayesian paper, \citet{Luo-Lian-Tian-2012} develop a
hierarchical model to estimate the parameters of conditional quantile
functions with random effects. The authors do so by adopting an Asymmetric
Laplace (AL) distribution for the residual errors and suitable prior
distributions for the parameters. However, directly using the AL
distribution does not yield tractable conditional densities for all of the
parameters and hence a combination of Metropolis-Hastings (MH) and Gibbs
sampling is required for model estimation. The use of the MH algorithm may
require tuning at each quantile. To overcome this limitation,
\citet{Luo-Lian-Tian-2012} also present a full Gibbs sampling algorithm that
utilizes the normal-exponential mixture representation of the AL
distribution. This mixture representation is also followed in our work, with
important computational improvements.

Finally, for discrete panel data, recent work by \cite{Bartolucci-Nigro-2010}
introduces a quadratic exponential model for binary panel data and utilizes a
conditional likelihood approach, which is computationally simpler than
previous classical estimators. Bayesian approaches to binary panel data
models include work by \cite{Albert-Chib-1996}, \cite{Chib-Carlin-1999},
\cite{Chib-Jeliazkov-2006}, and \cite{Burda-Harding-2013}. These work
influence the estimation methods designed in our quantile approach to binary
panel data.
\end{sloppypar}

This paper contributes to the three literatures by extending the various
methodologies to a hierarchical Bayesian quantile regression model for binary
longitudinal data and proposing a Markov chain Monte Carlo (MCMC) algorithm
to estimate the model. The model handles both common (fixed) and
individual-specific (random) parameters (commonly referred to mixed effects
in statistics). The algorithm implements a blocking procedure that is
computationally efficient and the distributions involved allow for
straightforward calculations of covariate effects. The framework is
implemented in two empirical applications. The first application examines
female labor force participation, which has been heavily studied in panel
form. The topic became of particular interest in the state dependence versus
heterogeneity debate \citep{Heckman-1981Hetero}. We revisit this question and
implement our panel quantile approach, which has been otherwise unexplored
for this topic. The results offer new insights regarding the determinants of
female labor force participation and how the ages of children have different
effects across the quantiles and utility scale. The findings suggest that
policy should be focused on women's transitions into the labor force after
child birth and the few years after.

The second application considers the probability of home ownership during the
Great Recession. Micro-level empirical analyses on individuals moving into
and out of housing markets are lacking in the recent literature. Past studies
include \cite{Carliner-1974} and \cite{Poirier-1977}, but the recent housing
crisis offers a new opportunity to reevaluate the topic. Furthermore, a full
quantile analysis of home ownership is yet to be explored. Since home
ownership is a choice that requires years of planning, individual
characteristics may range drastically across the latent utility scale. The
analysis presented in this paper controls for multivariate heterogeneity in
individuals and wealth, and investigates the determinants of home ownership,
state dependence in home ownership, and how the shock to housing markets
affected these items. The results provide an understanding as to how
individuals of particular demographics and socioeconomic status fared during
the collapse of the housing market.

The rest of the paper is organized as follows. Section~\ref{sec:QRandALD}
reviews quantile regression and the AL distribution, Section~\ref{sec:Model}
introduces the quantile regression model for binary longitudinal data,
presents a simulation study, and discusses methods for covariate effects.
Section~\ref{sec:Applications} considers the two applications and concluding
remarks are offered in Section~\ref{sec:Conclusion}.

\section{Quantile Regression and Asymmetric Laplace Distribution}\label{sec:QRandALD}
The $p$-{th} quantile of a random variable $Y$  is the value $y_{0}$ such
that the probability that $Y$ will be less than $y_{0}$ equals $p \in (0,1)$.
Mathematically, if $Q(\cdot)$ denotes the inverse of the cumulative
distribution function (\emph{cdf}) of $Y$, the $p$-{th} quantile is defined
as
\begin{equation*}
Q_{Y}(p) \equiv F_{Y}^{-1}(p)= \mathrm{inf} \{y_{0}: F(y_{0})\geq p\}.
\end{equation*}
Quantile regression implements the idea of quantiles within the regression
framework with $Q(\cdot)$ modified to denote the inverse \emph{cdf} of the
dependent variable given the covariates. The objective is to estimate
\emph{conditional quantile functions} and to this purpose, regression
quantiles are estimated by minimizing the quantile objective function which
is a sum of asymmetrically weighted absolute residuals.

\begin{figure*}[!t]
  \centerline{
    \mbox{\includegraphics[width=3.00in, height=2.00in]{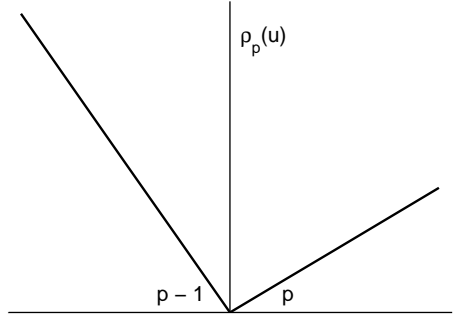}}
    }
\caption{Quantile regression check function} \label{fig:CheckFunction}
\end{figure*}

To formally explain the quantile regression problem, consider the following
linear model,
\begin{equation}
y_{i} = x'_{i} \beta_{p} + \varepsilon_{i}, \qquad \mathrm{with}
\qquad Q_{\varepsilon_{i}}(p|x_{i}) = 0,
\label{eq:linearModel}
\end{equation}
where $y_{i}$ is a scalar response variable, $x_{i}$ is a $k \times 1$ vector
of covariates, $\beta_{p}$ is a $k \times 1$ vector of unknown parameters
that depend on quantile $p$, and $\varepsilon_{i}$ is the error term such
that its $p$-th quantile equals zero. Henceforth, we will drop the subscript
$p$ for notational simplicity. In classical econometrics, the error
$\varepsilon$ does not (or is not assumed to) follow any distribution and
estimation requires minimizing the following objective function,
\begin{equation}
\min_{\beta \in \mathbf{R}^{k}} \bigg[\;  \sum_{i:y_{i} < x_{i}'\beta} (1 - p)\; |y_{i} - x_{i}'\beta| \;\;
      + \sum_{i:y_{i} \geq x_{i}'\beta}  p \; |y_{i} - x_{i}'\beta|  \;\; \bigg].
      \label{eq:objfunc}
\end{equation}
The minimizer $\hat{\beta}$ gives the $p$-th regression quantile and the
estimated conditional quantile function is obtained as $ \hat{y}_{i} = x'_{i}
\hat{\beta}$. Alternatively, the objective function \eqref{eq:objfunc} can be
written as a sum of piecewise linear or check functions as follows,
\begin{equation*}
\min_{\beta \in \mathbf{R}^{k}} \sum_{i=1}^{n}
\rho_{p}(y_{i} - x'_{i}\beta),
\label{eq:quantileCheckFunction}
\end{equation*}
where $\rho_{p}(u) = u \cdot (p - I(u<0))$ and $I(\cdot)$ is an indicator
function, which equals 1 if the condition inside the parenthesis is true and
0 otherwise. The check function, as seen in Figure~\ref{fig:CheckFunction},
is not differentiable at the origin. Hence, classical econometrics relies on
computational techniques to estimate quantile regression models. Such
computational methods include the simplex algorithm
\citep{Dantzig-1963,Dantzig-Thapa-1997,Dantzig-Thapa-2003,Barrodale-Roberts-1973,Koenker-dOrey-1987},
the interior point algorithm
\citep{Karmarkar-1984,Mehrotra-1992,Portnoy-Koenker-1997}, the smoothing
algorithm \citep{Madsen-Nielsen-1993, Chen-2007}, and metaheuristic
algorithms \citep{Rahman-2013}.

In contrast to classical quantile regression, Bayesian quantile regression
assumes that the error follows an AL distribution because the AL \emph{pdf}
contains the quantile loss function \eqref{eq:quantileCheckFunction} in its
exponent. This facilitates the construction of a working likelihood, required
for Bayesian analysis. Maximizing an AL likelihood is equivalent to
minimizing the quantile objective function
\citep{Koenker-Machado-1999,Yu-Moyeed-2001}. A random variable $Y$ follows an
AL distribution if its probability density function (\emph{pdf}) is given by:
\begin{equation}
f(y|\mu,\sigma,p) = \frac{p(1-p)}{\sigma} \exp \bigg[ -\rho_{p}
\bigg(\frac{y-\mu}{\sigma}  \bigg)  \bigg],
\label{eq:pdfLaplace}
\end{equation}
where $\rho_{p}(\cdot)$ is the check function as defined earlier, $ -\infty <
\mu < \infty$ is the location parameter,  $\sigma>0$ is the scale parameter,
and $0<p<1$ is the skewness parameter \citep{Kotz-etal-2001,Yu-Zhang-2005}.
The mean and variance of $Y$ with \emph{pdf} \eqref{eq:pdfLaplace} are
\begin{equation*}
E(Y) = \mu + \frac{\sigma(1 - 2p)}{p(1-p)} \quad \textrm{and} \quad
V(Y) = \frac{\sigma^{2}(1-2p+2p^{2})} {p^{2}(1-p)^{2}}.
\end{equation*}
If $\mu=0$ and $\sigma=1$, then both mean and variance depend only on $p$ and
hence are fixed for a given value of $p$.

The Bayesian approach to quantile regression for binary data assumes that
$\varepsilon \sim AL (0,1,p)$. Here, the variance is constant to serve as a
normalization for identification, typical in probit and logit models
\citep{Poirier-Ruud-1988,Koop-Poirier-1993,Jeliazkov-Rahman-2012}. However,
working directly with the AL distribution is not conducive to constructing a
Gibbs sampler and hence the normal-exponential mixture of the AL distribution
is often employed \citep{Kozumi-Kobayashi-2011}. Several recent papers have
utilized the mixture representation, including \citet{Ji-etal-2012} for
Bayesian model selection in binary and Tobit quantile regression,
\citet{Luo-Lian-Tian-2012} for estimating linear longitudinal data models,
and \citet{Rahman-2016} for estimating ordinal quantile regression models. We
also exploit the normal-exponential mixture representation of the AL
distribution to derive the estimation algorithm for quantile regression in
binary longitudinal data settings.

\section{The Quantile Regression Model for Binary Longitudinal Data}\label{sec:Model}
This section presents the quantile regression model for binary longitudinal
data (QBLD) and an estimation algorithm to fit the model. The performance of
the proposed algorithm is illustrated in a simulation study. The last part of
this section considers methods for model comparison and covariate effects.

\subsection{The Model}
The proposed model looks at quantiles of binary longitudinal data expressed
as a function of covariates with common effects and individual-specific
effects. The individual-specific effects offer additional flexibility in that
both intercept and slope heterogeneity can be captured, which are important
to avoid biases in the parameter estimates. The QBLD model can be
conveniently expressed in the latent variable formulation
\citep{Albert-Chib-1993} as follows,
\begin{equation}
\begin{split}
z_{it}  & =  x'_{it} \beta  + s'_{it}\alpha_{i} + \varepsilon_{it}, \hspace{0.5in}
\forall \; i=1, \cdots, n, \;\; t = 1, \cdots, T_{i},
\\
y_{it}  & =  \left\{ \begin{array}{ll}
1 & \textrm{if} \;  z_{it} > 0,\\
0 & \textrm{otherwise},
\end {array} \right.
\end{split}
\label{eq:BLDmodel}
\end{equation}
where the latent variable $z_{it}$ denotes the value of $z$ at the $t$-th
time period on the $i$-th individual, $x'_{it}$ is a $1 \times k$ vector of
explanatory variables, $\beta$ is $k \times 1$ vector of common parameters,
$s'_{it}$ is a $1 \times l$ vector of covariates that have
individual-specific effects, $\alpha_{i}$ is an $l \times 1$ vector of
individual-specific parameters, and $\varepsilon_{it}$ is the error term
assumed to be independently distributed as $AL(0,1,p)$ with
$Q_{\varepsilon_{it}}(p|x_{it},\alpha_{i}) = 0$. This implies that the
conditional density of $z_{it}|\alpha_{i}$ is an $AL( x'_{it}\beta + s'_{it}
\alpha_{i},1,p)$ for $i=1,\cdots,n$, and $t=1,\cdots,T_{i}$, with
$Q_{z_{it}}(p|x_{it},\alpha_{i}) = x'_{it}\beta + s'_{it} \alpha_{i}$. Note
that $s_{it}$ may contain a constant for intercept heterogeneity, as well as
other covariates (which are often a subset of those in $x_{it}$) to account
for slope heterogeneity of those variables. The variable $z_{it}$ is
unobserved and represents the latent utility associated with the observed
binary choice $y_{it}$. The latent variable formulation serves as a
convenient tool in the estimation process \citep{Albert-Chib-1993}.
Furthermore, latent utility underlies the interpretation of the results at
the various quantiles.

While working directly with the AL density is an option, the resulting
posterior will not yield the full set of tractable conditional distributions
necessary for a Gibbs sampler. Thus, we utilize the normal-exponential
mixture representation of the AL distribution, presented in
\citet{Kozumi-Kobayashi-2011}, and express the error as follows,
\begin{equation}
\varepsilon_{it} = \theta w_{it} + \tau \sqrt{w_{it}} \, u_{it},  \hspace{0.75in}
\forall \; i=1,\cdots,n; \; t = 1, \cdots, T_{i},
\label{eq:normal-exp}
\end{equation}
where $u_{it} \sim N(0,1)$ is mutually independent of $w_{it}
\sim\mathcal{E}(1)$ with $\mathcal{E}$ representing an exponential
distribution and the constants $\theta = \frac{1-2p}{p(1-p)}$ and $\tau =
\sqrt{\frac{2}{p(1-p)}}$. The mixture representation gives access to the
appealing properties of the normal distribution.

Longitudinal data models often involve a moderately large amount of data, so
it is important to take advantage of any opportunity to reduce the
computational burden. One such trick is to stack the model for each
individual $i$ \citep{Hendricks-Koenker-Poirier}. We define
$z_{i}=(z_{i1},\cdots,z_{i T_{i}})'$, $X_{i} = (x'_{i1},\cdots,x'_{i
T_{i}})'$, $S_{i} = (s'_{i1},\cdots,s'_{i T_{i}})'$, $w_{i} =
(w_{i1},\cdots,w_{i T_{i}})'$, $D_{\tau \sqrt{w_{i}}} = \mathrm{diag}(\tau
\sqrt{w_{i1}},\cdots,\tau \sqrt{w_{i T_{i}}})$, and $u_{i}=(u_{i1},\cdots,
u_{i T_{i}})'$. Building on equations~\eqref{eq:BLDmodel} and
\eqref{eq:normal-exp}, the resulting hierarchical model can be written as,
\begin{equation}
\begin{split}
z_{i}  & =  X_{i} \beta  + S_{i}\alpha_{i} + \theta w_{i} + D_{\tau\sqrt{w_{i}}} \; u_{i},
\\
y_{it}  & =  \left\{ \begin{array}{ll}
1 & \textrm{if} \;  z_{it} > 0,\\
0 & \textrm{otherwise},
\end {array} \right.
\\
\alpha_{i}|\varphi^{2}  & \sim N_{l}(0, \varphi^{2}I_{l}),
\hspace{0.6in} w_{it} \sim \mathcal{E}(1), \hspace{0.6in} u_{it} \sim N(0,1),
\\
\beta  & \sim N_{k}(\beta_{0}, B_{0}), \quad
\varphi^{2} \sim IG(c_{1}/2, d_{1}/2),
\end{split}
\label{eq:HierarchicalModel}
\end{equation}
where we assume that $\alpha_{i}$ are identically distributed as a normal
distribution. The last row represents the prior distributions with $N$ and
$IG$ denoting the normal and inverse-gamma distributions, respectively. Here,
we note that the form of the prior distribution on $\beta$ holds a penalty
interpretation on the quantile loss function \citep{Koenker-2004}. A normal
prior on $\beta$ implies a $\ell_{2}$ penalty and has been used in
\citet{Yuan-Yin-2010}, and \citet{Luo-Lian-Tian-2012}. One may also employ a
Laplace prior distribution on $\beta$ that imposes $\ell_{1}$ penalization,
as used in several articles such as \citet{Alhamzawi-Ali-2018}. While
\cite{Alhamzawi-Ali-2018} also work with quantile regression for discrete
panel data (ordered, in particular), our work contributes by considering
multivariate heterogeneity (not just intercept heterogeneity), and
introducing computational improvements outlined below.

By Bayes' theorem, we express the ``complete joint posterior'' density as
proportional to the product of likelihood function and the prior
distributions as follows,
\begin{equation}
\begin{split}
\pi(\beta,\alpha,w,z,\varphi^{2}|y) & \propto
\bigg\{ \prod_{i=1}^{n} f(y_{i}|z_{i},\beta,\alpha_{i},w_{i},\varphi^{2})
\pi(z_{i}|\beta,\alpha_{i},w_{i}) \pi(w_{i}) \pi(\alpha_{i}|\varphi^{2})
\bigg\}
\pi(\beta) \pi(\varphi^{2}),  \\
& \propto  \bigg\{ \prod_{i=1}^{n} \bigg[ \prod_{t=1}^{T_{i}} f(y_{it}|z_{it}) \bigg]
\pi(z_{i}|\beta,\alpha_{i},w_{i}) \pi(w_{i}) \pi(\alpha_{i}|\varphi^{2})
\bigg\}
\pi(\beta) \pi(\varphi^{2}),
\end{split}
\label{eq:jointPostBayesThm}
\end{equation}
where the first line uses independence between prior distributions and second
line follows from the fact that given $z_{it}$, the observed $y_{it}$ is
independent of all parameters because the second line of
\eqref{eq:HierarchicalModel} determines $y_{it}$ given $z_{it}$ with
probability 1. Substituting the distribution of the variables associated with
the likelihood and the prior distributions in \eqref{eq:jointPostBayesThm}
yields the following expression,
\begin{eqnarray}
&& \pi(\beta,\alpha,w,z,\varphi^{2}|y) \propto
 \bigg\{  \prod_{i=1}^{n}\prod_{t=1}^{T_{i}}
\Big[I(z_{it}>0) I(y_{it}=1) + I(z_{it} \leq 0) I(y_{it}=0) \Big] \bigg\}
\nonumber \\
&& \hspace{0.1in} \times
\exp\Big[- \frac{1}{2} \sum_{i=1}^{n} \Big\{(z_{i} - X_{i} \beta - S_{i}\alpha_{i}
- \theta w_{i})' \, D_{\tau \sqrt{w_{i}}}^{-2}
\, (z_{i} - X_{i} \beta - S_{i}\alpha_{i} - \theta w_{i}) \Big\} \Big]
\nonumber\\
&& \hspace{0.1in} \times \bigg\{ \prod_{i=1}^{n}
|D_{\tau \sqrt{w_{i}}}^{2}|^{-\frac{1}{2}}\bigg\} \times
\exp\bigg(- \sum_{i=1}^{n} \sum_{t=1}^{T_{i}} w_{it}
\bigg) \big(2 \pi \varphi^{2}\big)^{-\frac{nl}{2}}
\exp\Big[ -\frac{1}{2\varphi^{2}} \sum_{i=1}^{n}\alpha'_{i} \alpha_{i}
\Big]  \label{eq:jointPosterior} \\
&& \hspace{0.1in} \times (2\pi)^{-\frac{k}{2}} |B_{0}|^{-\frac{1}{2}} \exp\Big[ -\frac{1}{2} (\beta - \beta_{0})'
    B_{0}^{-1} (\beta - \beta_{0})  \Big] \times (\varphi^{2})^{-(\frac{c_{1}}{2} + 1)}
    \exp\Big[ - \frac{d_{1}}{2 \varphi^{2}} \Big]. \nonumber
\end{eqnarray}

The joint posterior density \eqref{eq:jointPosterior} does not have a
tractable form, and thus simulation techniques are necessary for estimation.
Bayesian methods are increasing in popularity \citep{Poirier2006}, and this
paper takes the approach for a couple of reasons. First, with discrete panel
data, working with the likelihood function is complicated because it is
analytically intractable. The inclusion of individual-specific effects makes
matters worse. Second, while numerical simulation methods are available for
discrete panel data, they are often slow and difficult to implement
\citep{Burda-Harding-2013}. The availability of a full set of conditional
distributions (which are outlined below) makes Gibbs sampling an attractive
option that will be simpler to implement, both conceptually and
computationally.

We can derive the conditional posteriors of the parameters and latent
variables by a straightforward extension of the estimation technique for the
linear mixed-effects model presented in \citet{Luo-Lian-Tian-2012}. This is
presented as Algorithm~\ref{alg:Non-block} in \ref{app:A}, which shows the
conditional posterior distributions for the parameters and latent variables
necessary for a Gibbs sampler. While this Gibbs sampler is straightforward,
there is potential for poor mixing properties due to correlation between
$(\beta, \alpha_i)$ and $(z_i,\alpha_i)$. The correlation often arises
because the variables corresponding to the parameters in $\alpha_i$ are often
a subset of those in $x_{it}$. Thus, by conditioning these items on one
another, the mixing of the Markov chain will be slow.

\begin{table*}[t!]
\begin{algorithm}[Blocked Sampling]
\label{alg:Block} \rule{\textwidth}{0.5pt} \small{
\begin{enumerate}[itemsep=1ex]
\item Sample $(\beta,z_i)$ in one block. The objects $(\beta,z_i)$ are sampled
in the following two sub-steps.
\begin{enumerate}[leftmargin=3ex]
    \item Let $\Omega_{i} =
         \left(\varphi^{2} S_{i} S'_{i} + D_{\tau \sqrt{w_{i}}}^{2}\right)$.
         Sample $\beta$ marginally of $\alpha$ from
         $\beta|z,w,\varphi^{2}$ $\sim$  $N(\tilde{\beta}, \tilde{B})$, where,
         \begin{equation*}
         \tilde{B}^{-1} = \bigg(\sum_{i=1}^{n}
         X'_{i} \Omega_{i}^{-1}   X_{i}
         + B_{0}^{-1} \bigg) \enskip
         \mathrm{and} \enskip
         \tilde{\beta} = \tilde{B}\left(\sum_{i=1}^{n} X'_{i} \Omega_{i}^{-1}
         (z_{i} - \theta w_{i}) + B_{0}^{-1}\beta_{0} \right).
         \end{equation*}
    \item Sample the vector $z_{i}|y_{i},\beta,w_{i},\varphi^{2} \sim
         TMVN_{B_{i}}(X_{i}\beta + \theta w_{i}, \Omega_{i})$ for all
         $i=1,\cdots,n$, where $B_{i} = (B_{i1} \times B_{i2} \times
         \ldots \times B_{iT_{i}})$ and $B_{it}$ is the interval $(0,\infty)$ if
         $y_{it}=1$ and the interval $(-\infty,0]$ if $y_{it}=0$. This is
         done by sampling $z_{i}$ at the $j$-th pass of the MCMC
         iteration using a series of conditional posterior distribution
         as follows:
         \begin{equation*}
         z_{it}^{j}|z_{i1}^{j},\cdots,z_{i(t-1)}^{j},z_{i(t+1)}^{j},\cdots,
         z_{iT_{i}}^{j} \sim  TN_{B_{it}}(\mu_{t|-t}, \Sigma_{t|-t}),
         \hspace{0.5in} \mathrm{for} \; t=1,\cdots,T_{i},
         \end{equation*}
         where $TN$ denotes a truncated normal distribution.
         The terms $\mu_{t|-t}$ and $\Sigma_{t|-t}$ are the conditional mean and
         variance, respectively, and are defined as,
         \begin{eqnarray*}
         \mu_{t|-t} & = & x'_{it}\beta + \theta w_{it} + \Sigma_{t,-t}
         \Sigma_{-t,-t}^{-1} \big( z_{i,-t}^{j} - (X_{i}\beta +
         \theta w_{i})_{-t} \big), \\
         \Sigma_{t|-t} & = & \Sigma_{t,t} - \Sigma_{t,-t}
         \Sigma_{-t,-t}^{-1}\Sigma_{-t,t},
         \end{eqnarray*}
         where $z_{i,-t}^{j}=(z_{i1}^{j}, \cdots,z_{i(t-1)}^{j},z_{i(t+1)}^{j-1},
         \cdots,z_{iT_{i}}^{j-1})$, $(X_{i}\beta + \theta w_{i})_{-t}$ is a
         column vector with $t$-th element removed, $\Sigma_{t,t}$ denotes
         the $(t,t)$-th element of $\Omega_{i}$, $\Sigma_{t,-t}$ denotes
         the $t$-th row of $\Omega_{i}$ with element in the $t$-th column
         removed and $\Sigma_{-t,-t}$ is the $\Omega_{i}$ matrix with
         $t$-th row and $t$-th column removed.
\end{enumerate}
\item    Sample $\alpha_{i}|z,\beta,w,\varphi^{2}$ $\sim$  $N(\tilde{a}, \tilde{A})$
         for $i=1,\cdots,n$, where,
         \begin{equation*}
         \tilde{A}^{-1} = \left(S'_{i} \, D^{-2}_{\tau \sqrt{w_{i}}} \, S_{i}
         + \frac{1}{\varphi^{2}}  I_{l}\right)
         \quad \mathrm{and} \quad
         \tilde{a} = \tilde{A} \left(S'_{i} D^{-2}_{\tau \sqrt{w_{i}}} \,
         \big(z_{i} - X_{i} \beta - \theta w_{i} \big)   \right).
         \end{equation*}
\item    Sample $w_{it}|z_{it},\beta,\alpha_{i}$ $\sim$  $GIG \,
         (0.5, \tilde{\lambda}_{it}, \tilde{\eta})$ for $i=1,\cdots,n$ and
         $t=1,\cdots,T_{i}$, where,
         \begin{equation*}
         \tilde{\lambda}_{it} = \bigg( \frac{ z_{it} - x'_{it}\beta - s'_{it}
         \alpha_{i}}{\tau} \bigg)^{2} \quad \mathrm{and} \quad \tilde{\eta}
         = \bigg(\frac{\theta^{2}}{\tau^{2}} + 2 \bigg).
         \end{equation*}
\item    Sample $\varphi^{2}|\alpha \sim IG(\tilde{c}_{1}/2, \tilde{d}_{1}/2)$, where
         $\tilde{c}_{1} = \Big( nl + c_{1} \Big)$ and
         $\tilde{d}_{1} = \Big( \displaystyle \sum_{i=1}^{n} \alpha_{i}'\alpha_{i}
         + d_{1} \Big)$.\\
\end{enumerate}}
\rule{\textwidth}{0.5pt}
\end{algorithm}
\end{table*}

To avoid this issue, we develop an alternative algorithm which jointly
samples $(\beta,z_i)$ in one block within the Gibbs sampler. This blocked
approach significantly improves the mixing properties of the Markov chain.
The success of these blocking techniques can be found in \cite{Liu94},
\citet{Chib-Carlin-1999}, and \cite{Chib-Jeliazkov-2006}. The details of our
blocked sampler are described in Algorithm~\ref{alg:Block}.\footnote{The
derivation of the conditional posterior densities are presented in Appendix
B.} In particular, $\beta$ is sampled marginally of $\alpha_i$ from a
multivariate normal distribution. The latent variable $z_{i}$ is sampled
marginally of $\alpha_{i}$ from a truncated multivariate normal distribution
denoted by $TMVN_{B_{i}}$, where $B_{i}$ is the truncation region given by
$B_{i} = (B_{i1} \times B_{i2} \times \ldots \times B_{iT_{i}})$ such that
$B_{it}$ is the interval $(0,\infty)$ if $y_{it}=1$ and the interval
$(-\infty,0]$ if $y_{it}=0$. To draw from a truncated multivariate normal
distribution, we utilize the method proposed in \citet{Geweke-1991}. This
involves drawing from a series of conditional posteriors which are univariate
truncated normal distributions. Previous work using this approach include
\citet{Chib-Greenberg-1998} and \citet{Chib-Carlin-1999}. The random effects
parameter $\alpha_i$ is sampled conditionally on $\beta,z_i$ from another
multivariate normal distribution. The variance parameter $\varphi^{2}$ is
sampled from an inverse-gamma distribution and finally the latent weight $w$
is sampled element-wise from a generalized inverse Gaussian (GIG)
distribution \citep{Dagpunar-1988,Dagpunar-1989,Devroye-2014}.

\begin{sloppypar}
We end this section with a cautionary note on sampling from a truncated
multivariate normal distribution, with the hope that it will be useful to
researchers on quantile regression. In our algorithm above, we sample $z_{i}$
from a $TMVN_{B_{i}}(X_{i}\beta + \theta w_{i}, \Omega_{i})$ using a series
of conditional posteriors which are univariate truncated normal
distributions. This method is distinctly different and should not be confused
with sampling from a recursively characterized truncation region typically
related to the Geweke-Hajivassiliou-Keane (GHK) estimator \citep{Geweke-1991,
BorschSupan-Hajivassiliou-1994, Keane-1994,
Hajivassiliou-McFadden-1998}.\footnote{In the latter scenario, the model
$z_{i} \sim N(X_{i}\beta + \theta w_{i}, \Omega_{i}) $ can be written as
$z_{i} = X_{i}\beta + \theta w_{i} + L_{i} \eta_{i}$, where $L_{i}$ is a
lower triangular Cholesky factor of $\Omega_{i}$ such that
$L_{i}L'_{i}=\Omega_{i}$. To be general, let the lower and upper truncation
vectors for $z_{i}$ be $a_{i}=(a_{i1}, \ldots, a_{iT_{i}})$ and
$b_{i}=(b_{i1}, \ldots, b_{iT_{i}})$, respectively. Then the random variable
$\eta_{it}$ is sampled from $TN\big(0,1, (a_{it} - x'_{it}\beta - \theta
w_{it} - \sum_{j=1}^{t-1} l_{tj} \eta_{ij})/l_{tt}, (b_{it} - x'_{it}\beta -
\theta w_{it} - \sum_{j=1}^{t-1} l_{tj} \eta_{ij})/l_{tt} \big)$, where
$l_{tj}$ are the elements of $L_{i}$. This is a recursively characterized
truncation region, since the range of $\eta_{it}$ depends on the draw of
$\eta_{ij}$ for $j=1,\ldots,t-1$. The vector $z_{i}$ can be obtained by
substituting the recursively drawn $\eta_{i}$ into $z_{i}=X_{i}\beta + \theta
w_{i} + L\eta_{i}$. However, the draws so obtained are not the same as
drawing $z_{i}$ from a multivariate normal distribution truncated to the
region $a_{i} < z_{i} < b_{i}$.} The difference between the two samplers have
been exhibited in \citet{Breslaw-1994} and carefully discussed in
\citet{Jeliazkov-Lee-2010}.
\end{sloppypar}

\subsection{Simulation Study}\label{subsec:SimStudies}
This subsection evaluates the performance of the algorithm in a simulation
study, where the data are generated from a model that has common effects and
individual-specific effects in both the intercept and slopes. We estimate the
quantile regression model for binary longitudinal data (QBLD) using our
proposed blocked sampler (Algorithm~\ref{alg:Block}) and the non-blocked
sampler (Algorithm~\ref{alg:Non-block}).

The data are simulated from the model $z_{it} = x'_{it}\beta +
s'_{it}\alpha_{i} + \varepsilon_{it}$ where $t = 1, \ldots, 10$ and $i = 1,
\ldots, 500$. For the parameters and covariates: $\beta = (-5, 6, 4)'$,
$\alpha_{i} \sim N(0_{2}, I_{2})$, $x'_{it}=(1, x_{2it}, x_{3it})$ with
$x_{2it} \sim U(0,1)$ and $x_{3it}\sim U(0,1)$, $s'_{it} = (1, s_{2it})$ with
$s_{2it} \sim U(0,1)$. The error is generated from a standard AL
distribution, $\varepsilon_{it} \sim AL(0,1,p)$ for $p=0.25, 0.5, 0.75$.
Here, the notation $U(0,1)$ denotes a standard uniform distribution. The
binary response variable $y_{it}$ is constructed by assigning 1 to all
positive values of $z_{it}$ and 0 to all negative values of $z_{it}$. Since
the values generated from an AL distribution are different at each quantile,
the number of 0s and 1s are also different at each quantile. In the
simulation, the number of observations corresponding to 0s and 1s for the
25th, 50th and 75th quantiles are $(1566,3444)$, $(2588, 2412)$ and $(3536,
1464)$, respectively.

\begin{table}[t!]
\centering \footnotesize \setlength{\tabcolsep}{4pt} \setlength{\extrarowheight}{2pt}
\setlength\arrayrulewidth{1pt}
\caption{Posterior means (\textsc{mean}), standard deviations (\textsc{std})
and inefficiency factors (\textsc{if}) of the parameters in the simulation
study from the QBLD model. The first panel presents results from
Algorithm~\ref{alg:Block} and the second panel presents results
from Algorithm~\ref{alg:Non-block}. }
\begin{tabular}{c rrr rrr rrr rrr r }
\toprule
& & \multicolumn{11}{c}{Blocked Sampling} &  \\
\cmidrule{3-13}
& & \multicolumn{3}{c}{\textsc{25th quantile}} & & \multicolumn{3}{c}{\textsc{50th quantile}}
& & \multicolumn{3}{c}{\textsc{75th quantile}} &  \\
\cmidrule{3-5} \cmidrule{7-9}  \cmidrule{11-13}
      & &  \textsc{mean} & \textsc{std} & \textsc{if}
      & &  \textsc{mean} & \textsc{std} & \textsc{if}
      & &  \textsc{mean} & \textsc{std} & \textsc{if} &   \\
\midrule
$\beta_{1}$     & & $-5.33$  & $0.22$  & $ 4.55$  & & $-5.06$  & $0.18$  & $ 4.09$
                & & $-5.08$  & $0.24$  & $ 4.10$  &  \\
$\beta_{2}$     & & $ 6.16$  & $0.28$  & $ 4.38$  & & $ 5.96$  & $0.22$  & $ 3.87$
                & & $ 6.16$  & $0.27$  & $ 4.11$  &  \\
$\beta_{3}$     & & $ 4.34$  & $0.24$  & $ 3.86$  & & $ 3.88$  & $0.19$  & $ 3.66$
                & & $ 3.88$  & $0.23$  & $ 3.21$  &  \\
$\varphi^{2}$   & & $ 0.95$  & $0.16$  & $ 4.68$  & & $ 0.66$  & $0.11$  & $ 4.60$
                & & $ 0.81$  & $0.15$  & $ 4.93$  &  \\
\midrule
& & \multicolumn{11}{c}{Non-blocked Sampling} &  \\
\cmidrule{3-13}
& & \multicolumn{3}{c}{\textsc{25th quantile}} & & \multicolumn{3}{c}{\textsc{50th quantile}}
& & \multicolumn{3}{c}{\textsc{75th quantile}} &  \\
\cmidrule{3-5} \cmidrule{7-9}  \cmidrule{11-13}
          & &  \textsc{mean} & \textsc{std} & \textsc{if}
          & &  \textsc{mean} & \textsc{std} & \textsc{if}
          & &  \textsc{mean} & \textsc{std} & \textsc{if} &   \\
\midrule
$\beta_{1}$     & & $-5.32$  & $0.22$  & $ 5.94$  & & $-5.05$ & $0.20$ & $ 6.90$
                & & $-5.07$  & $0.23$  & $ 6.63$  &  \\
$\beta_{2}$     & & $ 6.15$  & $0.27$  & $ 6.05$  & & $ 5.95$ & $0.23$ & $ 6.57$
                & & $ 6.15$  & $0.26$  & $ 6.69$  &  \\
$\beta_{3}$     & & $ 4.35$  & $0.24$  & $ 5.52$  & & $ 3.88$ & $0.20$ & $ 5.40$
                & & $ 3.88$  & $0.23$  & $ 5.34$  &  \\
$\varphi^{2}$   & & $ 0.95$  & $0.16$  & $ 5.58$  & & $ 0.66$ & $0.11$ & $ 5.26$
                & & $ 0.81$  & $0.14$  & $ 6.15$  &  \\
\bottomrule
\end{tabular}
\label{Table:SimResult1}
\end{table}

The posterior estimates of the model parameters are based on the generated
data and the following independent prior distributions: $\beta \sim
N(0_{k},10 I_{k})$, and $\varphi^{2} \sim IG(10/2, 9/2)$.
Table~\ref{Table:SimResult1} reports the posterior means, standard deviations
and inefficiency factors calculated from $12,000$ MCMC iterations after a
burn-in of $3,000$ iterations. The inefficiency factors are calculated using
the batch-means method discussed in \citet{Greenberg-2012}. The simulation
exercise was repeated for various covariates, sample sizes, common and
individual-specific parameters, and the results do not change from this
baseline case; hence they are not presented.

The posterior mean for regression coefficients for both the samplers (blocked
and non-blocked methods) are near the true values, $\beta = (-5, 6, 4)'$.
Additionally, the standard deviations are small. Across each quantile, the
number of 0s and 1s varies, and the samplers perform well in each case.
Furthermore, starting the algorithm at different values appears
inconsequential, which is a benefit of the full Gibbs sampler.

Turning attention to the differences between the two algorithms, it is clear
that the inefficiency factors from the blocked algorithm are much lower,
suggesting better sampling performance and a nice mixing of the Markov chain.
The advantages of the blocking procedure are more apparent from the
autocorrelation in the MCMC draws at different lags.
Table~\ref{Table:Autocorr} presents the autocorrelation in MCMC draws at
lag~1, lag~5, and lag~10. Looking at lag~10, the autocorrelation for the
$\beta$s are between $0.25-0.43$ in the blocked algorithm, which is nearly
half of $0.55-0.73$, obtained from the non-blocked sampler. Recall that in
our data generation process, we did not make the covariates in $s_{it}$ a
subset of those in $x_{it}$. Whereas in real-data exercises, it is typical
for $s_{it}$ to be a subset. Therefore, we expect the benefits of the blocked
sampler to be even more pronounced in real data settings.

\begin{table}[t!]
\centering \footnotesize \setlength{\tabcolsep}{4pt} \setlength{\extrarowheight}{2pt}
\setlength\arrayrulewidth{1pt}
\caption{Autocorrelation in MCMC draws at Lag~1, Lag~5 and Lag~10.}
\begin{tabular}{c rrr rrr rrr rrr r }
\toprule
& & \multicolumn{11}{c}{Blocked Sampling} &  \\
\cmidrule{3-13}
& & \multicolumn{3}{c}{\textsc{25th quantile}} & & \multicolumn{3}{c}{\textsc{50th quantile}}
& & \multicolumn{3}{c}{\textsc{75th quantile}} &  \\
\cmidrule{3-5} \cmidrule{7-9}  \cmidrule{11-13}
      & &  Lag~1 & Lag~5 & Lag~10
      & &  Lag~1 & Lag~5 & Lag~10
      & &  Lag~1 & Lag~5 & Lag~10 &   \\
\midrule
$\beta_{1}$     & & $0.86$  & $0.59$  & $0.41$  & & $0.85$  & $0.54$  & $0.35$
                & & $0.88$  & $0.61$  & $0.41$  &  \\
$\beta_{2}$     & & $0.89$  & $0.61$  & $0.43$  & & $0.87$  & $0.53$  & $0.34$
                & & $0.89$  & $0.60$  & $0.39$  &  \\
$\beta_{3}$     & & $0.86$  & $0.50$  & $0.31$  & & $0.83$  & $0.44$  & $0.25$
                & & $0.84$  & $0.45$  & $0.23$  &  \\
$\varphi^{2}$   & & $0.93$  & $0.73$  & $0.54$  & & $0.92$  & $0.70$  & $0.51$
                & & $0.93$  & $0.75$  & $0.58$  &  \\
\midrule
& & \multicolumn{11}{c}{Non-blocked Sampling} &  \\
\cmidrule{3-13}
& & \multicolumn{3}{c}{\textsc{25th quantile}} & & \multicolumn{3}{c}{\textsc{50th quantile}}
& & \multicolumn{3}{c}{\textsc{75th quantile}} &  \\
\cmidrule{3-5} \cmidrule{7-9}  \cmidrule{11-13}
          & &  Lag~1 & Lag~5 & Lag~10
          & &  Lag~1 & Lag~5 & Lag~10
          & &  Lag~1 & Lag~5 & Lag~10 &   \\
\midrule
$\beta_{1}$     & & $0.96$  & $0.84$  & $0.71$  & & $0.97$  & $0.85$  & $0.73$
                & & $0.97$  & $0.85$  & $0.71$  &  \\
$\beta_{2}$     & & $0.96$  & $0.81$  & $0.68$  & & $0.96$  & $0.82$  & $0.68$
                & & $0.96$  & $0.80$  & $0.65$  &  \\
$\beta_{3}$     & & $0.95$  & $0.77$  & $0.61$  & & $0.95$  & $0.77$  & $0.60$
                & & $0.94$  & $0.75$  & $0.55$  &  \\
$\varphi^{2}$   & & $0.92$  & $0.76$  & $0.63$  & & $0.92$  & $0.74$  & $0.59$
                & & $0.93$  & $0.79$  & $0.68$  &  \\
\bottomrule
\end{tabular}
\label{Table:Autocorr}
\end{table}

\begin{figure*}[!t]
	\centerline{
		\mbox{\includegraphics[width=6.75in, height=3.25in]{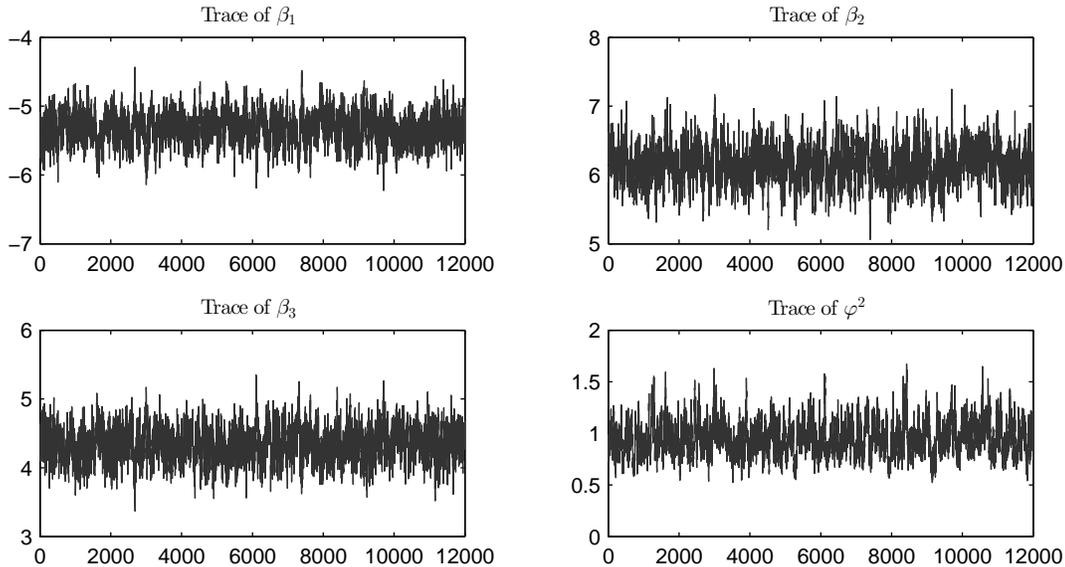}} }
	\caption{Trace plots of the MCMC draws at the 25th quantile
            from Algorithm~\ref{alg:Block}.}
\label{fig:traceplots}
\end{figure*}

Finally, Figure~\ref{fig:traceplots} presents the trace plots of the
parameters at the 25th quantile for the blocked algorithm, which graphically
demonstrate the appealing sampling. Given the computational efficiency with
the blocking procedure, it is our preferred way for estimating QBLD models
and will be used in the subsequent real data applications.

\subsection{Additional Considerations}\label{subsec:AdditionalConsiderations}
In this section, we briefly discuss methods for model comparison and
computation of covariate effects. For model comparison, we follow standard
techniques for longitudinal data models. Specifically, in the application
sections we provide the log-likelihood, conditional AIC
\citep{Greven-Kneib-2010}, and conditional BIC \citep{Delattre-etal-2014}.
This is a bit unusual for a Bayesian analysis, however, we want the results
in our empirical applications to align with the classical work on the topics,
such as \cite{Bartolucci-Farcomeni-2012}. Thus, we follow the approaches so
as to allow for better comparisons and cross references.

For covariate effects, in general terms, we are interested in the average
difference in the implied probabilities between the case when $x_{1it}$ is
set to the value $x_{1it}^{\dag }$ and $x_{1it}^{\ddag}$ . Given the values
of the other covariates denoted $x_{-1it}, s_{it}$ and those of the model
parameters $\theta$, one can obtain the probabilities $\Pr
(y_{it}=1|x_{1it}^{\dag},x_{-1it},s_{it},\theta )$ and $\Pr
(y_{it}=1|x_{1it}^{\ddag },x_{-1it},s_{it},\theta )$. Following from
\cite{Jeliazkov-etal-2008} and \cite{Jeliazkov-Vossmeyer-2018}, if one is
interested in the distribution of the difference $\{\Pr
(y_{it}=1|x_{1it}^{\dag })-\Pr (y_{it}=1|x_{1it}^{\ddag })\} $ marginalized
over $\{x_{-1it}, s_{it}\}$ and $\theta $ given the data $y$, a practical
procedure is to marginalize out the covariates using their empirical
distribution, while the parameters are integrated out with respect to their
posterior distribution. Formally, the goal is to obtain a sample of draws
from the distribution,
\begin{equation*}
\begin{split}
& \{ \Pr (y_{it}=1|x_{1it}^{\dag })-\Pr (y_{it}=1|x_{1it}^{\ddag }) \}  \\
& = \int \{\Pr
(y_{it}=1|x_{1it}^{\dag },x_{-1it},s_{it},\theta )-\Pr (y_{it}=1|x_{1it}^{\ddag
},x_{-1it},s_{it},\theta )\} \; \\
& \qquad \times \pi (x_{-1it},s_{it}) \,
\pi(\theta |y) \; d(x_{-1it},s_{it}) \; d\theta.
\end{split}
\end{equation*}
The computation of these probabilities is straightforward because the
differences between the probabilities of success is related to differences in
AL \emph{cdf}, marginalized over $\{x_{-1it}, s_{it}\}$ and the posterior
distribution of $\theta$. Also, the procedure handles uncertainty stemming
from the sample and estimation strategy. This approach is demonstrated in
each of the following applications.

\section{Applications}\label{sec:Applications}

\subsection{Female Labor Force Participation} \label{sec:WomenLabor}

Modeling female labor force participation has been an important area of work
in the economics and econometric literature for decades. The list of work is
vast, but a partial list includes \citet{Heckman-Macurdy-1980},
\citet{Heckman-Macurdy-1982}, \citet{Mroz-1987}, \citet{Hyslop-1999},
\citet{Arellano-Carrasco-2003}, \citet{Chib-Jeliazkov-2006},
\cite{Kordas-2006}, \citet{Carro-2007}, \citet{Bartolucci-Nigro-2010}, and
\citet{Eckstein-Lifshitz-2011}.

Within the literature, several pertinent questions have been analyzed
including the relationship between participation and age, education,
fertility, and permanent and transitory incomes. However, serial persistence
in the decision to participate and its two competing theories --
heterogeneity and state dependence -- have been of substantive interest.
Heterogeneity implies that females may differ in terms of certain unmeasured
variables that affect their probability of labor force participation. If
heterogeneity is not properly controlled, then past decisions may appear
significant to current decisions leading to what is called spurious state
dependence. In contrast, pure state dependence implies that dynamic effects
of past participation genuinely affect current employment decisions.
Consideration of heterogeneity and state dependence is important in modeling
female labor force participation and can have economic implications as
discussed in \citet{Heckman-1981Hetero}, \citet{Heckman-1981DPD} and
\citet[pp. 261-270]{Hsiao-2014}. We re-examine the above mentioned aspects
using our proposed Bayesian quantile regression model for binary longitudinal
data. To our knowledge, this is the first attempt to analyze female labor
force participation within a longitudinal quantile framework. So, what can we
learn from a panel quantile approach? Of particular interest are the impacts
of infants and children across the various quantiles. Understanding the
differential effects across the latent utility scale can help shape female
labor force policies, such as maternity leave and child care.

Before proceeding forward, we draw attention to \cite{Kordas-2006} who
evaluated female labor force participation using cross sectional data and
smoothed binary regression quantiles. His results offer interesting insights
across the quantiles, which further motivate our application and extension to
transitions into and out of the labor force in the panel setting. We also
follow his interpretation where the latent utility differential between
working and not working may be interpreted as a ``propensity'' or
``willingness-to-participate'' (WTP) index.

\begin{table}[!b]
\centering \footnotesize \setlength{\tabcolsep}{6pt} \setlength{\extrarowheight}{1pt}
\setlength\arrayrulewidth{1pt}
\caption{Sample characteristics of the female labor force participation data  --
The first panel presents the mean/proportion and standard deviations (in parenthesis) of
the variables in the full and the sub-samples. The second panel displays
the column percentages for the number of years worked and the third panel
(i.e., last row) presents the number of observations in the full
and the sub-samples.}
\label{Table:FLFPDataSummary}
\begin{tabular}{p{2.3cm}  c  >{\raggedleft}p{1.2cm} >{\raggedleft}p{1.5cm}
                >{\raggedleft}p{1.5cm} >{\raggedleft}p{1.7cm}
                >{\raggedleft}p{1.6cm} p{1.8cm} <{\raggedleft} }
\toprule
                & & Full Sample               & Employed 7 Years
                  & Employed 0 Years          & Single Transition from Work
                  & Single Transition to Work & Multiple Transitions     \\
                & & (1) & (2) & (3) & (4) & (5) & (6) \\
\midrule
Age            & & $29.55$   & $30.44$   & $29.18$   & $29.21$  & $29.23$   & $28.68$  \\
               & & ($4.61$)  & ($4.34$)  & ($4.51$)  & ($4.77$) & ($4.62$)  & ($4.73$) \\
Education      & & $13.14$   & $13.33$   & $12.68$   & $13.20$  & $13.01$   & $13.08$  \\
               & & ($2.06$)  & ($1.98$)  & ($2.15$)  & ($2.13$) & ($2.19$)  & ($2.05$) \\
Child 1-2      & & $0.31$    & $0.22$    & $0.46$    & $0.31$   & $0.34$    & $0.38$   \\
               & & ($0.53$)  & ($0.45$)  & ($0.60$)  & ($0.53$) & ($0.57$)  & ($0.57$) \\
Child 3-5      & & $0.37$    & $0.27$    & $0.56$    & $0.32$   & $0.50$    & $0.42$   \\
               & & ($0.57$)  & ($0.49$)  & ($0.65$)  & ($0.54$) & ($0.65$)  & ($0.60$) \\
Child 6-13     & & $0.75$    & $0.71$    & $0.92$    & $0.55$   & $0.99$    & $0.74$   \\
               & & ($0.92$)  & ($0.87$)  & ($1.00$)  & ($0.81$) & ($1.03$)  & ($0.94$) \\
Child 14-      & & $0.32$    & $0.39$    & $0.31$    & $0.29$   & $0.26$    & $0.26$   \\
               & & ($0.67$)  & ($0.72$)  & ($0.71$)  & ($0.69$) & ($0.61$)  & ($0.60$) \\
Black          & & $0.24$    & $0.27$    & $0.26$    & $0.19$   & $0.21$    & $0.22$   \\
               & & ($0.43$)  & ($0.44$)  & ($0.44$)  & ($0.39$) & ($0.40$)  & ($0.41$) \\
Income/$10,000$& & $ 3.04$   & $ 2.82$   & $ 3.81$   & $ 3.43$  & $ 2.99$   & $ 2.96$   \\
               & & ($2.60$)  & ($1.82$)  & ($5.28$)  & ($3.14$) & ($2.04$)  & ($1.89$) \\
Fertility      & & $ 0.07$   & $ 0.04$   & $ 0.08$   & $ 0.10$  & $ 0.05$   & $ 0.09$   \\
               & & ($0.25$)  & ($0.21$)  & ($0.28$)  & ($0.29$) & ($ 0.22$) & ($0.28$)  \\
\midrule
Years worked   & &    &    &     &    &      &  \\
0              & & $ 10.30$   & $ -   $   & $ 100 $   & $  -    $  & $  -    $   & $  -    $   \\
1              & & $  5.33$   & $ -   $   & $ -   $   & $ 20.00 $  & $  9.03 $   & $  8.25 $   \\
2              & & $  6.29$   & $ -   $   & $ -   $   & $  7.86 $  & $ 12.26 $   & $ 14.39 $   \\
3              & & $  6.36$   & $ -   $   & $ -   $   & $ 12.14 $  & $ 10.97 $   & $ 13.68 $   \\
4              & & $  8.64$   & $ -   $   & $ -   $   & $ 11.43 $  & $ 17.42 $   & $ 19.34 $   \\
5              & & $  9.34$   & $ -   $   & $ -   $   & $ 13.57 $  & $ 23.23 $   & $ 18.87 $   \\
6              & & $ 13.76$   & $ -   $   & $ -   $   & $ 35.00 $  & $ 27.10 $   & $ 25.47 $   \\
7              & & $ 39.97$   & $ 100 $   & $ -   $   & $ -     $  & $ -     $   & $ -     $   \\
\midrule
Observations   & & $ 1446 $   & $ 578 $   & $ 149 $    & $ 140 $   & $ 155 $     & $ 424 $   \\
\bottomrule
\end{tabular}
\end{table}

The data for this study are taken from \citet{Bartolucci-Farcomeni-2012},
which were originally extracted from the Panel Study of Income Dynamics
(PSID) conducted by the University of Michigan. The data consist of a sample
of $n=1446$ females who were followed for the period 1987 to 1993 with
respect to their employment status and a host of demographic and
socio-economic variables. The dependent variable in the model is
\emph{employment} status ($=1$ if the individual is employed, $=0$ otherwise)
and the covariates include \emph{age} (in 1986), \emph{education} (number of
years of schooling), \emph{child 1-2} (number of children aged 1 to 2,
referred to the previous year), \emph{child 3-5}, \emph{child 6-13},
\emph{child 14-}, \emph{Black} (indicator for Black race), \emph{income} of
the husband (in US dollars, referred to the previous year), and
\emph{fertility} (indicator variable for birth of a child in a certain year).
Lagged employment status is also included as a covariate to examine state
dependence of female labor force participation decision.

Table~\ref{Table:FLFPDataSummary} presents summary statistics for the
variables. The presentation of the table follows from \cite{Hyslop-1999},
where statistics are broken up into subgroups of women that have worked 0
years, 7 years, or transitioned during the period. As one can see from the
table, the average age in the sample is roughly 30, about 40\% of the sample
is employed throughout the entire period, 10\% are not in the labor force
throughout the entire period, 20\% transition into or out of the labor force
once, and 30\% transition multiple times. Looking closely at the different
variables for children, there is a decent amount of variation across the
subgroups. For mothers who are employed 0 years, the average values for
\textit{child 1-2} and \textit{child 3-5} are 0.46 and 0.56, respectively.
These numbers are more than double compared to that of mothers who are
employed for all the 7 years. Further, as children age (\textit{child 6-13})
more mothers have a single transition to work. While these differences
demonstrate some observed heterogeneity, unobserved heterogeneity still plays
a role, which motivates further analysis. Particularly, a quantile setting
will reveal information not available in the raw observed data by utilizing
the latent scale as the willingness-to-participate index.

The data are modeled following equations~\eqref{eq:BLDmodel}
and~\eqref{eq:normal-exp} and the model (QBLD) is specified with a random
intercept (i.e., $s_{it}$ only includes a constant). We also estimate the
probit model for binary longitudinal data (PBLD) using the algorithm
presented in \cite{Koop-Poirier-Tobias-2007} and \citet{Greenberg-2012} and
identical priors for relevant parameters. The results for the QBLD and PBLD
models are presented in Table~\ref{Table:FLFPResult} and are based on data
for the years 1988-1993, since using a lagged dependent variable drops
information for the year 1987. The reported posterior estimates are based on
12,000 MCMC draws after a burn-in of 3,000 draws and the following priors on
the parameters: $\beta \sim N(0_{k}, 10 I_{k})$ and $\varphi^{2} \sim
IG(10/2, 9/2)$. Table~\ref{Table:FLFPResult} presents the posterior means,
standard deviations, and inefficiency factors at the 25th, 50th, and 75th
quantiles, and for the binary probit model. Furthermore, the log-likelihood,
conditional AIC \citep{Greven-Kneib-2010} and conditional BIC
\citep{Delattre-etal-2014} are available for each model.

\begin{table}[!t]
\centering \footnotesize \setlength{\tabcolsep}{4pt} \setlength{\extrarowheight}{2pt}
\setlength\arrayrulewidth{1pt}
\caption{Results from the female labor force participation study --
Posterior means (\textsc{mean}), standard deviations (\textsc{std})
and inefficiency factors (\textsc{if}) of the parameters from the QBLD and PBLD models are provided.}
\begin{tabular}{l rrr rrr rrr rrr rrrr r  }
\toprule
& & \multicolumn{11}{c}{\textsc{QBLD}} && \multicolumn{3}{c}{}  \\
\cmidrule{3-13}
& & \multicolumn{3}{c}{\textsc{25th quantile}} & & \multicolumn{3}{c}{\textsc{50th quantile}}
& & \multicolumn{3}{c}{\textsc{75th quantile}} & &   \multicolumn{3}{c}{\textsc{PBLD}}  \\
\cmidrule{3-5} \cmidrule{7-9}  \cmidrule{11-13} \cmidrule{15-17}
      & &  \textsc{mean} & \textsc{std} & \textsc{if}
      & &  \textsc{mean} & \textsc{std} & \textsc{if}
      & &  \textsc{mean} & \textsc{std} & \textsc{if}
      & &  \textsc{mean} & \textsc{std} & \textsc{if} &  \\
\midrule
Intercept       & & $-3.11$  & $0.21$  & $4.59$  & & $-0.31$ & $0.18$ & $4.45$
                & & $ 1.35$  & $0.23$  & $4.79$  & & $-0.08$ & $0.07$ & $2.79$ \\
$\mathrm{Age}^{\dagger}$
                & & $ 0.03$  & $0.01$  & $2.27$  & & $ 0.01$ & $0.01$ & $2.38$
                & & $-0.01$  & $0.02$  & $2.72$  & & $ 0.01$ & $0.01$ & $1.57$ \\
$(\mathrm{Age^{\dagger}})^{2}/100$
                & & $-0.23$  & $0.26$  & $1.96$  & & $-0.19$ & $0.25$ & $2.06$
                & & $-0.13$  & $0.33$  & $2.59$  & & $-0.08$ & $0.10$ & $1.45$ \\
$\mathrm{Education}^{\dagger}$
                & & $ 0.17$  & $0.03$  & $2.29$  & & $ 0.21$ & $0.03$ & $2.57$
                & & $ 0.28$  & $0.05$  & $3.18$  & & $ 0.08$ & $0.01$ & $1.59$ \\
Child 1-2       & & $-0.22$  & $0.11$  & $2.68$  & & $-0.28$ & $0.11$ & $2.84$
                & & $-0.38$  & $0.13$  & $2.97$  & & $-0.12$ & $0.04$ & $1.67$ \\
Child 3-5       & & $-0.55$  & $0.10$  & $2.89$  & & $-0.52$ & $0.10$ & $3.22$
                & & $-0.56$  & $0.12$  & $2.91$  & & $-0.21$ & $0.04$ & $1.74$ \\
Child 6-13      & & $-0.17$  & $0.07$  & $2.39$  & & $-0.18$ & $0.07$ & $2.59$
                & & $-0.18$  & $0.08$  & $2.95$  & & $-0.07$ & $0.02$ & $1.58$ \\
Child 14-       & & $-0.05$  & $0.10$  & $2.66$  & & $-0.02$ & $0.10$ & $2.89$
                & & $-0.01$  & $0.13$  & $3.22$  & & $-0.01$ & $0.04$ & $1.71$ \\
Black           & & $ 0.20$  & $0.15$  & $2.02$  & & $ 0.24$ & $0.15$ & $2.24$
                & & $ 0.26$  & $0.19$  & $2.69$  & & $ 0.09$ & $0.06$ & $1.53$ \\
$\mathrm{Income}^{\dagger}/10,000$
                & & $-0.13$  & $0.03$  & $3.03$  & & $-0.14$ & $0.02$ & $3.00$
                & & $-0.18$  & $0.03$  & $3.52$  & & $-0.05$ & $0.01$ & $1.95$ \\
Fertility       & & $-1.91$  & $0.20$  & $2.71$  & & $-2.06$ & $0.20$ & $2.90$
                & & $-2.60$  & $0.33$  & $3.85$  & & $-0.72$ & $0.07$ & $1.67$ \\
Lag Employment  & & $ 4.89$  & $0.16$  & $3.75$  & & $ 3.88$ & $0.13$ & $4.47$
                & & $ 6.71$  & $0.20$  & $5.24$  & & $ 1.49$ & $0.05$ & $3.34$ \\
$\varphi^{2}$   & & $ 1.42$  & $0.35$  & $6.36$  & & $ 1.39$ & $0.33$ & $6.16$
                & & $ 2.12$  & $0.50$  & $7.12$  & & $ 0.33$ & $0.05$ & $4.97$ \\

\midrule
Log-likelihood
& & \multicolumn{3}{c}{$-3115.72$} & & \multicolumn{3}{c}{$-3127.38$}
& & \multicolumn{3}{c}{$-3146.68$} & & \multicolumn{3}{c}{$-2887.91$}  \\
AIC
& & \multicolumn{3}{c}{$6257.45$}  & & \multicolumn{3}{c}{$6280.77$}
& & \multicolumn{3}{c}{$6319.36$}  & & \multicolumn{3}{c}{$5801.82$}  \\
BIC
& & \multicolumn{3}{c}{$6354.82$}  & & \multicolumn{3}{c}{$6378.14$}
& & \multicolumn{3}{c}{$6416.74$}  & & \multicolumn{3}{c}{$5899.20$}  \\
\bottomrule
\multicolumn{17}{l}{\textsuperscript{$\dagger$}\footnotesize{denotes variable
minus the sample average.}}
\end{tabular}
\label{Table:FLFPResult}
\end{table}

First, note that across the quantiles the inefficiency factors are low,
implying a nice mixing of the Markov chain. These results, which were
demonstrated in the simulation study, hold in empirical applications as well.
Next, if we consider each quantile as corresponding to a different
likelihood, then the 25th quantile has the lowest conditional AIC and
conditional BIC. This result is not surprising since the unconditional
probability of participation is around 70\% in the sample. Our result also
finds support in \cite{Kordas-2006}, where he reports that the 30th
conditional quantile would be the one most
efficiently estimable. 

The results for the education variable are positive, statistically different
from zero, and show various incremental differences across the quantiles.
Education is found to have stronger effects in the upper part of the latent
index, which is expected since these are women who have a high utility for
working and thus have obtained the requisite education. Regarding the state
dependence versus heterogeneity debate, we find that employment is serially
positively correlated, which is a consequence of state dependence. The effect
gets incrementally larger as one moves up the latent utility scale. While we
are controlling for individual heterogeneity with the random intercept, we
still find evidence of state dependence. This result agrees with
\cite{Bartolucci-Farcomeni-2012}, who investigate the question with a latent
class model. Other papers that find empirical evidence of strong state
dependence effects include \cite{Heckman-1981Hetero}, \cite{Hyslop-1999}, and
\cite{Chib-Jeliazkov-2006}.

To further understand the results, covariate effects are computed for several
variables for the 3 quantiles and the PBLD model. The covariate effect
calculations follow from Section~\ref{subsec:AdditionalConsiderations} and
the results are displayed in Table~\ref{Table:FLFPCovEff}. Note that the 50th
quantile results are similar to that of the PBLD, which is to be expected.
The covariate effect for education is calculated on the restricted sample of
individuals with a high school degree (12 years of schooling). The effect
that is computed is 4 additional years of schooling, implying a college
degree. The effect for income is a discrete change by \$10,000, the effect
for children is increasing the count by one, and for fertility it is a
discrete change to the indicator variable.

\begin{table}[!t]
\centering \footnotesize \setlength{\tabcolsep}{6pt} \setlength{\extrarowheight}{2pt}
\setlength\arrayrulewidth{1pt}
\caption{Covariate effects in the female labor force participation study.}
\begin{tabular}{l rrr rrr r }
\toprule
& & \multicolumn{3}{c}{\textsc{QBLD}} && \multicolumn{1}{c}{}  \\
\cmidrule{3-5}
                        & &  \textsc{25th} & \textsc{50th}
                        &  \textsc{75th} & & \textsc{PBLD}   &  \\
\midrule
Education     & & $ 0.0523$  & $ 0.0711$  & $ 0.0633$  & & $ 0.0698$  & \\
Child 1-2     & & $-0.0160$  & $-0.0212$  & $-0.0206$  & & $-0.0254$  & \\
Child 3-5     & & $-0.0415$  & $-0.0397$  & $-0.0302$  & & $-0.0430$  & \\
Child 6-13    & & $-0.0123$  & $-0.0133$  & $-0.0098$  & & $-0.0146$  & \\
Income        & & $-0.0095$  & $-0.0102$  & $-0.0097$  & & $-0.0105$  & \\
Fertility     & & $-0.1672$  & $-0.1747$  & $-0.1335$  & & $-0.1627$  & \\
\bottomrule
\end{tabular}
\label{Table:FLFPCovEff}
\end{table}

The results show that the birth of a child in that year (\textit{fertility}),
reduces the probability that a woman works by 16.7 percentage points at the
25th quantile, 17.4 percentage points at the 50th quantile, and 13.3
percentage points at the 75th quantile. For individuals in the lower part of
the latent index, having children ages 1-2 impacts their employment decision
less than those at the upper quantiles. Perhaps, women with a low utility for
working are less impacted by infants and toddlers because it is often a
desire to stay home with the child for a few years. Whereas, women with a
high utility for working face negative impacts because of the desire to enter
the work force.

The most pronounced negative effect of children occurs when the child is ages
3-5. Often women temporarily exit the work force until children are ready for
pre-school and this result provides evidence of the difficulty mothers faces
re-entering the work force after a several year leave of absence
\citep{Drange-Rege-2013}. The finding is interesting from a policy
standpoint. If policy is focused on increasing participation, offering more
support in the years when the child is likely not breastfeeding but before
kindergarten would be beneficial.

The covariate effect of a college degree is 5.2 to 7.1 percentage points
across the quantiles, while husband's income is approximately $-1$ percentage
point across the quantiles. Thus, a college degree increases the probability
a woman works by about 6 percentage points, whereas an increase in family
income only decreases the probability by 1 percentage point for every
\$10,000. While many of these results align with existing findings, the
behavior in the high and low quantiles presents useful information, which was
otherwise unexplored in panel data.

\subsection{Home Ownership} \label{sec:Housing}

\begin{figure*}[!b]
  \centerline{
    \mbox{\includegraphics[width=6.75in, height=2.5in]{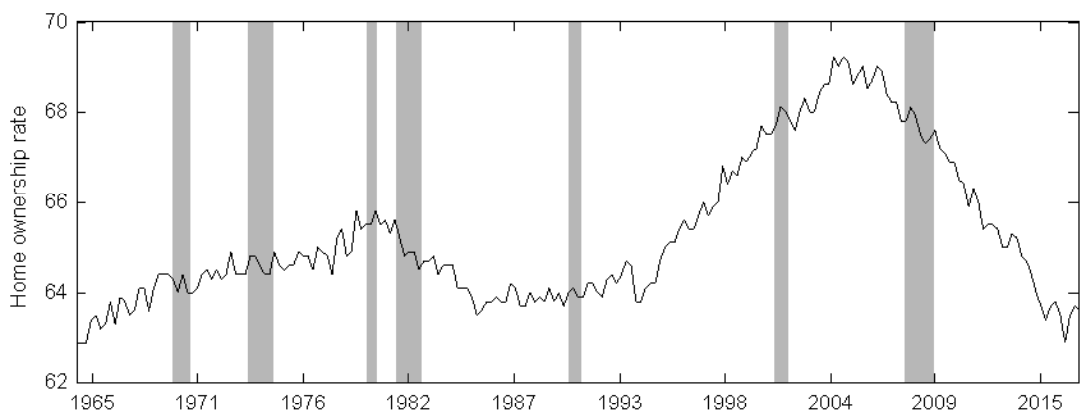}}
    }
\caption{Home ownership rates in the United States. Data taken from FRED,
provided by the Federal Reserve Bank of St. Louis. } \label{Fig:HomeOwn}
\end{figure*}

The recent financial crisis had major implications for home ownership in the
United States. Figure~\ref{Fig:HomeOwn} displays the home ownership rates for
the United States from the 1960s to 2017. These data were taken from the FRED
website provided by the Federal Reserve Bank of St. Louis. The rate of home
ownership rose in the late 1990s and early 2000s, but started to decline
after 2007. The determinants of home ownership was reviewed in the 1970s
\citep{Carliner-1974,Poirier-1977}. However, the recent crisis offers a
unique event and shock to housing markets to reevaluate this topic.

\begin{sloppypar}
The literature on home ownership has examined racial gaps
\citep{Charles-Hurst-2002, Turner-Smith-2009}, wealth accumulation and income
\citep{Turner-Luea-2009}, mobility and the labor market
\citep{Ferreira-Gyourko-Tracy-2010, Fairlie-2013}, and tax policy
\citep{Hilber-Turner-2014}. However, unlike the labor force context, state
dependence has only been lightly examined with regard to housing
tenure.\footnote{\cite{Chen-Ost-2005} control for state dependence in a study
of housing allowance in Sweden.} Given the large down payments and extensive
mortgage processes typical in home ownership, state dependence is likely to
be a key factor, as well as individual heterogeneity.

Furthermore, quantile analyses in the home ownership literature are lacking.
The quantiles represent degrees of \textit{willingness} or \textit{utility}
of owning a home. Owning a home in the United States usually requires an
individual to produce a large upfront investment, a promising credit history,
and a willingness to engage in 30 year mortgages, resulting in less
liquidity. Given these requirements, interest lies in how the determinants of
home ownership varies across the latent utility scale. Therefore, this paper
adds to the literature on the probability of home ownership by employing the
QBLD model. The approach has several advantages, namely that we can control
for multivariate heterogeneity, visit the state dependence versus
heterogeneity argument in the housing context, and analyze willingness of
home ownership across the quantiles.
\end{sloppypar}



The dataset is constructed from the Panel Study of Income Dynamics (PSID) and
consists of a balanced panel of 4092 individuals observed for the years 2001,
2003, 2005, 2007, 2009, 2011, and 2013. The sample is restricted to
individuals aged 25-65 who answered the relevant questions for the 7 years
and captures the period before, during, and after the Great Recession. The
dependent variable is defined as follows:
\begin{equation}
y_{it}   =  \left\{ \begin{array}{ll}
1 & \textrm{home owner}    \\
0 & \textrm{not a home owner},
\end{array} \right.
\label{eq:HomeOwner}
\end{equation}
for $i = 1, \ldots, 4092$ and $t = 2003,~2005,2007,~2009,~2011,~2013$ (2001
is dropped because it is a dynamic model). The covariates include
demographics, marital status, employment, job industry, health insurance,
education, socioeconomic status, lagged home ownership, and an indicator for
after the recession (2009-2013). The model includes a random intercept and a
random slope on an income-to-needs variable, which allows for individual
heterogeneity and heterogeneity in income. Heterogeneity in income is an
important control because a marginal increase in income could have a wide
range of effects on the probability of owning a home, where for some the
effect of income could be 0 (perhaps, those who own their home freehold, or
those who have no desire for ownership). Whereas, for others, increases in
income could go directly into home ownership utility.
Table~\ref{Table:HomeOwnDataSummary} presents summary statistics for the
variables. Once again, the presentation of the table follows from
\cite{Hyslop-1999}, where statistics are broken up into subgroups of people
that have always been home owners, never been home owners, or transitioned
during the period of interest.

\afterpage{
\begin{footnotesize}
\setstretch{1.20}    
\begin{longtable}{p{3.0cm}  c  >{\raggedleft}p{1.2cm} >{\raggedleft}p{1.5cm}
                >{\raggedleft}p{1.5cm} >{\raggedleft}p{1.8cm}
                >{\raggedleft}p{1.8cm} p{1.8cm} <{\raggedleft} }
\caption{Sample characteristics of the home ownership data  -- The first
panel presents the mean and standard deviations (in parenthesis) of the
continuous variables and proportions of the categorical variables in the full
and the sub-samples. The second panel displays the column percentages for the
number of years home is owned and the third panel (i.e., last row) presents
the number of observations in the full and the
sub-samples.}\label{Table:HomeOwnDataSummary}\\
\toprule
                & & Full Sample               & Owned 6 Years
                  & Owned 0 Years             & Single Transition from Ownership
                  & Single Transition to Ownership & Multiple Transitions     \\
\midrule
                & & (1) & (2) & (3) & (4) & (5) & (6) \\
\midrule
\endfirsthead
\multicolumn{8}{c}%
{\tablename\ \thetable\ -- \textit{Continued from previous page}} \\
\hline
\endhead
\hline
\endfoot
Age of Head (2003)    & & $45.74$   & $48.78$   & $42.31$   & $46.71$  & $38.91$   & $39.62$  \\
               & &($13.51$)  &($12.56$)  &($13.82$)  &($15.48$) &($11.63$)  &($12.29$) \\
No. Children   & & $ 0.78$   & $ 0.71$   & $ 0.82$   & $ 0.83$  & $ 0.85$   & $ 1.00$  \\
               & & ($1.15$)  & ($1.08$)  & ($1.25$)  & ($1.21$) & ($1.12$)  & ($1.27$) \\
Income/$10,000$& & $ 7.71$   & $ 9.66$   & $ 3.17$   & $ 6.54$  & $ 6.74$   & $ 6.82$   \\
               & &($10.53$)  &($11.18$)  & ($3.33$)  &($16.28$) & ($5.71$)  &($10.15$) \\
Inc-Needs Ratio
               & & $ 4.89$   & $ 6.07$   & $ 2.17$   & $ 4.25$  & $ 4.37$   & $ 4.23$   \\
               & & ($7.38$)  & ($7.51$)  & ($2.30$)  &($14.01$) & ($3.74$)  & ($5.66$)  \\
Net Wealth/$10,000$
               & & $23.05$   & $35.76$   & $ 1.98$   & $12.40$  & $ 8.20$   & $10.98$   \\
               & &($128.08$) &($167.12$) &($16.79$)  &($54.30$) &($32.70$)  &($45.75$) \\
Female         & & $0.23$    & $0.14$    & $0.49$    & $0.26$   & $0.22$    & $0.24$   \\
Married        & & $0.61$    & $0.78$    & $0.23$    & $0.49$   & $0.53$    & $0.48$   \\
Single         & & $0.15$    & $0.06$    & $0.42$    & $0.11$   & $0.20$    & $0.17$   \\
Below Bachelors&
               & $0.54$      & $0.52$    & $0.53$    & $0.58$   & $0.57$    & $0.57$   \\
Bachelors \& Above
               & & $0.27$    & $0.34$    & $0.11$    & $0.19$   & $0.26$    & $0.22$   \\
Job Cat1       & & $0.58$    & $0.61$    & $0.54$    & $0.61$   & $0.53$    & $0.54$   \\
Job Cat2       & & $0.11$    & $0.12$    & $0.07$    & $0.10$   & $0.12$    & $0.12$   \\
Job Cat3       & & $0.06$    & $0.06$    & $0.03$    & $0.06$   & $0.08$    & $0.07$   \\
Health Insurance
               & & $0.92$    & $0.97$    & $0.83$    & $0.90$   & $0.92$    & $0.87$   \\
Race-Black     & & $0.30$    & $0.20$    & $0.58$    & $0.32$   & $0.31$    & $0.35$   \\
Race-Others    & & $0.04$    & $0.04$    & $0.06$    & $0.05$   & $0.04$    & $0.05$   \\
Head Unemployed
               & & $0.06$    & $0.03$    & $0.12$    & $0.07$   & $0.06$    & $0.08$   \\
Head NLF       & & $0.23$    & $0.24$    & $0.28$    & $0.29$   & $0.11$    & $0.18$   \\
West           & & $0.19$    & $0.19$    & $0.18$    & $0.20$   & $0.19$    & $0.20$   \\
South          & & $0.41$    & $0.38$    & $0.44$    & $0.45$   & $0.45$    & $0.48$   \\
North-East     & & $0.14$    & $0.15$    & $0.14$    & $0.10$   & $0.16$    & $0.10$   \\
\hline
Years owned    & &    &    &     &    &      &  \\
0              & & $ 18.43$   & $ -   $   & $ 100 $   & $  -    $  & $  -    $   & $  -    $   \\
1              & & $  4.28$   & $ -   $   & $ -   $   & $ 18.10 $  & $ 14.89 $   & $ 17.34 $   \\
2              & & $  4.57$   & $ -   $   & $ -   $   & $ 18.40 $  & $ 14.10 $   & $ 21.39 $   \\
3              & & $  4.11$   & $ -   $   & $ -   $   & $ 19.94 $  & $ 14.63 $   & $ 13.87 $   \\
4              & & $  5.50$   & $ -   $   & $ -   $   & $ 20.55 $  & $ 23.67 $   & $ 19.94 $   \\
5              & & $  7.16$   & $ -   $   & $ -   $   & $ 23.01 $  & $ 32.71 $   & $ 27.46 $   \\
6              & & $ 55.96$   & $ 100 $   & $ -   $   & $ -     $  & $  -    $   & $  -    $   \\
\midrule
Observations   & & $ 4092 $   & $ 2290 $  & $  754 $   & $ 326    $  & $ 376 $     & $ 346 $   \\
\bottomrule
\end{longtable}
\end{footnotesize}
}

In the sample, about 56\% of individuals own a home across the entire sample
period, 18\% never own, and the remaining transition at least once. The age
of the head of the household is that in the year 2003. Job industry is
classified into four categories. \textit{JobCat1} is an indicator for jobs in
construction, manufacturing, agriculture, and wholesale. \textit{JobCat2} is
an indicator for jobs in business, finance, and real estate. \textit{JobCat3}
is an indicator for jobs in the military and public services. The omitted
category (\textit{JobCat4}) consists of jobs in professional and technical
services, entertainment and arts services, health care, and other. Education
is broken up into categories: less than high school (omitted), high school
degree or some college  (\textit{Below Bachelors}), and college or advanced
degree (\textit{Bachelors \& Above}). Race is broken up into white/asian
(omitted), \textit{black}, and \textit{other}. Marital status is discretized
into \textit{married}, \textit{single}, and divorced/widowed (omitted).
Region is discretized to \textit{west}, \textit{south}, \textit{northeast},
and midwest (omitted). We have two income measures, including
\textit{income-to-needs ratio} and \textit{net wealth}.\footnote{This measure
of net wealth excludes home equity and housing assets, so as to not conflate
with the outcome of interest.} We employ an inverse hyperbolic sine (IHS)
transformation for net wealth because it adjusts for skewness and retains
negative and 0 values, which is a common feature of data on net wealth
\citep{Friedline-etal-2015}.

The table demonstrates some drastic differences across the subgroups. As
expected, the ``owned 6 years'' group is older and wealthier than the others.
Families that transition tend to have more children, and a higher proportion
of females and singles are in the ``owned 0 years'' group. These differences
in the raw data motivate our question of interest -- with so much state
dependence in home ownership and heterogeneity among individuals and income,
what are the determinants of home ownership through an economic downturn? The
results should provide insights into discrepancies across subgroups of the
population and should better inform policy aiming to assist home owners
during downturns. Standard methods for investigating a binary panel dataset
of this sort do not capture the extensive heterogeneity problem, nor do they
offer quantile analyses, which highlights the usefulness of our approach.

The results for the home ownership application are presented in
Table~\ref{Table:HomeOwnResults}. Posterior means, standard deviations, and
inefficiency factors calculated using the batch-means method are presented
for the 25th, 50th, and 75th quantiles, as well as for the binary
longitudinal probit model (PBLD). The results are based on 12,000 MCMC draws
with a burn in of 3,000 draws.  The priors on the parameters are: $\beta \sim
N(0_{k}, 10 I_{k})$, and $\varphi^{2} \sim IG(10/2, 9/2)$.  As in the female
labor force application, the inefficiency factors are low, implying a nice
mixing of the Markov chain.

\begin{table}[!t]
\centering \footnotesize \setlength{\tabcolsep}{3.5pt} \setlength{\extrarowheight}{2pt}
\setlength\arrayrulewidth{1pt}
\caption{Posterior means (\textsc{mean}), standard deviations (\textsc{std})
and inefficiency factors (\textsc{if}) of the parameters in the QBLD model
and PBLD model for the home ownership application.}
\begin{tabular}{l rrr rrr rrr rrr rrrr r  }
\toprule
& & \multicolumn{11}{c}{\textsc{QBLD}} && \multicolumn{3}{c}{}  \\
\cmidrule{3-13}
& & \multicolumn{3}{c}{\textsc{25th quantile}} & & \multicolumn{3}{c}{\textsc{50th quantile}}
& & \multicolumn{3}{c}{\textsc{75th quantile}} & &   \multicolumn{3}{c}{\textsc{PBLD}}  \\
\cmidrule{3-5} \cmidrule{7-9}  \cmidrule{11-13} \cmidrule{15-17}
      & &  \textsc{mean} & \textsc{std} & \textsc{if}
      & &  \textsc{mean} & \textsc{std} & \textsc{if}
      & &  \textsc{mean} & \textsc{std} & \textsc{if}
      & &  \textsc{mean} & \textsc{std} & \textsc{if} &  \\
\midrule
Intercept        & & $-15.25$  & $0.85$  & $3.47$  & & $-9.26$ & $0.80$ & $4.36$
                 & & $ -4.17$  & $0.88$  & $3.47$  & & $-4.15$ & $0.31$ & $2.27$ \\
log Age of Head  & & $ 1.63$  & $0.21$  & $3.40$  & & $ 0.97$ & $0.20$ & $3.76$
                 & & $ 0.09$  & $0.23$  & $3.47$  & & $ 0.54$ & $0.08$ & $2.16$ \\
No. children     & & $ 0.14$  & $0.05$  & $3.46$  & & $ 0.18$ & $0.05$ & $4.32$
                 & & $ 0.22$  & $0.05$  & $3.65$  & & $ 0.08$ & $0.02$ & $2.13$ \\
Inc-Needs Ratio  & & $ 0.48$  & $0.03$  & $8.42$  & & $ 0.45$ & $0.03$ & $8.10$
                 & & $ 0.55$  & $0.04$  & $6.68$  & & $ 0.23$ & $0.01$ & $6.07$ \\
IHS Net Wealth   & & $ 0.25$  & $0.03$  & $4.05$  & & $ 0.32$ & $0.03$ & $4.56$
                 & & $ 0.41$  & $0.04$  & $5.11$  & & $ 0.10$ & $0.01$ & $2.60$ \\
Female           & & $ 0.95$  & $0.14$  & $3.21$  & & $ 0.82$ & $0.14$ & $3.84$
                 & & $ 0.59$  & $0.18$  & $3.79$  & & $ 0.25$ & $0.05$ & $1.90$ \\
Married          & & $ 2.28$  & $0.14$  & $3.42$  & & $ 2.18$ & $0.15$ & $4.05$
                 & & $ 1.70$  & $0.17$  & $4.35$  & & $ 0.74$ & $0.05$ & $2.05$ \\
Single           & & $ 0.32$  & $0.15$  & $3.37$  & & $ 0.17$ & $0.15$ & $4.13$
                 & & $-0.27$  & $0.17$  & $3.49$  & & $ 0.01$ & $0.05$ & $1.97$ \\
Below Bachelors  & & $ 0.17$  & $0.12$  & $3.71$  & & $ 0.28$ & $0.11$ & $3.80$
                 & & $ 0.35$  & $0.14$  & $3.89$  & & $ 0.11$ & $0.04$ & $2.02$ \\
Bachelors \& Above
                 & & $ 0.28$  & $0.18$  & $3.81$  & & $ 0.37$ & $0.16$ & $4.11$
                 & & $ 0.51$  & $0.20$  & $4.52$  & & $ 0.13$ & $0.06$ & $2.28$ \\
JobCat1          & & $ 0.31$  & $0.13$  & $4.23$  & & $ 0.39$ & $0.12$ & $4.34$
                 & & $ 0.55$  & $0.13$  & $4.07$  & & $ 0.16$ & $0.04$ & $2.20$ \\
JobCat2          & & $ 0.06$  & $0.20$  & $3.92$  & & $ 0.21$ & $0.19$ & $4.52$
                 & & $ 0.35$  & $0.21$  & $3.86$  & & $ 0.08$ & $0.07$ & $2.29$ \\
JobCat3          & & $ 0.03$  & $0.24$  & $3.54$  & & $ 0.08$ & $0.23$ & $3.82$
                 & & $ 0.11$  & $0.26$  & $3.74$  & & $ 0.01$ & $0.08$ & $2.04$ \\
Health Insurance & & $ 0.46$  & $0.16$  & $3.78$  & & $ 0.46$ & $0.15$ & $3.90$
                 & & $ 0.23$  & $0.19$  & $4.49$  & & $ 0.09$ & $0.05$ & $2.05$ \\
Race-Black       & & $-0.40$  & $0.12$  & $3.80$  & & $-0.54$ & $0.12$ & $3.55$
                 & & $-0.52$  & $0.14$  & $3.81$  & & $-0.18$ & $0.04$ & $1.96$ \\
Race-Others      & & $-0.15$  & $0.22$  & $3.43$  & & $-0.51$ & $0.21$ & $3.80$
                 & & $-0.85$  & $0.25$  & $4.20$  & & $-0.19$ & $0.07$ & $1.90$ \\
Head Unemployed  & & $-0.91$  & $0.18$  & $3.83$  & & $-0.89$ & $0.19$ & $4.27$
                 & & $-0.76$  & $0.23$  & $5.13$  & & $-0.23$ & $0.06$ & $2.14$ \\
Head NLF         & & $-0.40$  & $0.14$  & $4.03$  & & $-0.28$ & $0.14$ & $4.39$
                 & & $-0.06$  & $0.16$  & $4.42$  & & $-0.10$ & $0.05$ & $2.32$ \\
West             & & $-0.46$  & $0.15$  & $3.38$  & & $-0.49$ & $0.15$ & $3.78$
                 & & $-0.52$  & $0.18$  & $3.95$  & & $-0.16$ & $0.06$ & $1.98$ \\
South            & & $ 0.15$  & $0.13$  & $3.29$  & & $ 0.22$ & $0.13$ & $3.88$
                 & & $ 0.26$  & $0.14$  & $3.97$  & & $ 0.10$ & $0.05$ & $2.10$ \\
Northeast        & & $-0.28$  & $0.18$  & $3.44$  & & $-0.43$ & $0.17$ & $3.78$
                 & & $-0.59$  & $0.20$  & $3.84$  & & $-0.17$ & $0.06$ & $2.15$ \\
Post-Recession (PR)
                 & & $-1.32$  & $0.23$  & $6.22$  & & $-0.53$ & $0.13$ & $4.28$
                 & & $-0.44$  & $0.11$  & $2.91$  & & $-0.09$ & $0.04$ & $2.05$ \\
lag-Home Own     & & $ 7.46$  & $0.17$  & $5.93$  & & $ 5.90$ & $0.12$ & $4.25$
                 & & $ 9.55$  & $0.20$  & $7.04$  & & $ 2.19$ & $0.04$ & $2.15$ \\
PR*(lag-Home Own)
                 & & $ 1.47$  & $0.25$  & $5.99$  & & $ 0.72$ & $0.18$ & $4.94$
                 & & $ 0.76$  & $0.29$  & $8.53$  & & $ 0.11$ & $0.06$ & $2.60$ \\
$\varphi^{2}$    & & $ 0.13$  & $0.01$  & $9.88$  & & $ 0.11$ & $0.01$ & $8.67$
                 & & $ 0.16$  & $0.02$  & $8.94$  & & $ 0.04$ & $0.01$ & $8.71$ \\

\midrule
Log-likelihood
& & \multicolumn{3}{c}{$-5077.07$} & & \multicolumn{3}{c}{$-5030.12$}
& & \multicolumn{3}{c}{$-5085.64$} & & \multicolumn{3}{c}{$-4446.37$}  \\
AIC
& & \multicolumn{3}{c}{$10204.14$} & & \multicolumn{3}{c}{$10110.24$}
& & \multicolumn{3}{c}{$10221.27$} & & \multicolumn{3}{c}{$8942.73$}  \\
BIC
& & \multicolumn{3}{c}{$10421.69$} & & \multicolumn{3}{c}{$10327.79$}
& & \multicolumn{3}{c}{$10438.83$} & & \multicolumn{3}{c}{$9160.29$}  \\
\bottomrule
\end{tabular}
\label{Table:HomeOwnResults}
\end{table}

Many of the results agree with the existing literature. Income, education,
and being married all have a positive effect on home ownership
\citep{Turner-Smith-2009, Hilber-Turner-2014}. While these align with
intuition, new insights are offered across the quantiles for many of the
variables. Education, for instance, is not statistically different from zero
at the lower quantile. If one has a low utility for home ownership, education
will not impact that decision. Additionally, age of the head has a positive
impact on home ownership at the lower and median quantiles. However, for
those who have a high \textit{utility} for home ownership, age of the head is
not statistically different from 0. Number of children, on the other hand,
has a positive impact across the quantiles. Family growth seems to play a
role in owning a home.

The coefficient for female is positive which implies that females relative to
males are more in favor of home ownership. Given that housing was previously
thought of as a safe investment, this finding aligns with
\cite{Croson-Gneezy-2009}, who investigate gender differences in preferences
and find that women are more risk averse than men. Furthermore, relative to
divorced/widowed individuals, being single has a positive effect only at the
lower quantile. Interestingly, health insurance has a positive effect at the
lower and middle quantiles and is not statistically different from zero at
the higher willingness. Thus, if one has a high utility for home ownership,
potential costs related to health do not play into the the decision to invest
in a home. While \textit{race-black} is negative across the quantiles, which
is consistent with findings in \cite{Charles-Hurst-2002}, \textit{race-other}
is meaningful and negative only at the middle and upper quantiles. Thus,
policy interested in race disparities in home ownership, should focus on high
willingness individuals, because low willingness \textit{race-other}
individuals are not statistically different from whites.

The coefficient for Post-Recession (2009-2013) is negative across all of the
quantiles. This finding is expected given the major collapse in housing
markets. The state dependence variable (\textit{lag-Home Own}) is very large
and positive for all of the quantiles. Even with a shock to housing markets
and heterogeneity in the intercept and income controlled for, state
dependence is a key element of home ownership. Interestingly, the interaction
term between the state dependence variable and the post-recession indicator
has a credibility interval that includes 0 for the PBLD model, but is
positive across the quantiles. This finding is intriguing because the
positive state dependence effect offsets the negative effect from the
recession. Perhaps individuals who did not own a home prior to the recession
had trouble transitioning to ownership as a result of the tightened lending
and credit channels. This reasoning falls in line with the work of
\cite{Hilber-Turner-2014} in that mortgage policies can effect subgroups of
home owners, but not in aggregate. The aggregate finding in PBLD shows the
result is not statistically different from 0, but we find new results at the
quantiles.

Covariate effect calculations, which follow from the discussion in
Section~\ref{subsec:AdditionalConsiderations}, are computed for several
variables in both of the models, QBLD and PBLD. The results are displayed in
Table~\ref{Table:HOCovEff}, and show that being a female increases the
probability of home ownership by 2.9 to 1.6 percentage points, for the 25th
and 75th quantiles, respectively. The size of the effect is roughly halved at
the 75th quantile. This is useful for understanding the differences in
preferences between males and females, in particular, that at a higher
willingness, they are more similar than at a lower willingness. Similar
differing effects are found for the variable married, where being married
increases the probability of home ownership by 8.7 percentage points at the
25th quantile and 5.4 percentage points at the 75th quantile. Furthermore,
health insurance increases the probability of home ownership by 1.5
percentage points at a low willingness and 0.06 percentage points at the high
willingness (although the basic result at the 75th quantile was not different
from 0).

\begin{table}[!t]
\centering \footnotesize \setlength{\tabcolsep}{6pt} \setlength{\extrarowheight}{2pt}
\setlength\arrayrulewidth{1pt}
\caption{Covariate effects in the home ownership study. Age is increased by
10 years and the untransformed net wealth is increased by \$50,000.
The rest of the variable are indicators.}
\begin{tabular}{l rrr rrr r }
\toprule
& & \multicolumn{3}{c}{\textsc{QBLD}} && \multicolumn{1}{c}{}  \\
\cmidrule{3-5}
                        & &  \textsc{25th} & \textsc{50th}
                        &  \textsc{75th} & & \textsc{PBLD}   &  \\
\midrule
log Age of Head    & & $ 0.0111$  & $ 0.0076$  & $ 0.0006$  & & $ 0.0128$  & \\
IHS Net Wealth     & & $ 0.0120$  & $ 0.0179$  & $ 0.0203$  & & $ 0.0177$  & \\
Female             & & $ 0.0298$  & $ 0.0289$  & $ 0.0168$  & & $ 0.0264$  & \\
Married            & & $ 0.0879$  & $ 0.0890$  & $ 0.0548$  & & $ 0.0916$  & \\
Bachelors \& Above & & $ 0.0089$  & $ 0.0132$  & $ 0.0153$  & & $ 0.0144$  & \\
Health Insurance   & & $ 0.0159$  & $ 0.0165$  & $ 0.0063$  & & $ 0.0098$  & \\
Race-Black         & & $-0.0133$  & $-0.0193$  & $-0.0149$  & & $-0.0194$  & \\
Head Unemployed    & & $-0.0337$  & $-0.0326$  & $-0.0203$  & & $-0.0250$  & \\
\bottomrule
\end{tabular}
\label{Table:HOCovEff}
\end{table}

The aforementioned results find smaller effects at the higher willingness,
however, this is not the case for education and wealth. Wealth and education
have a greater impact for those with a high utility. Increasing net wealth by
\$50,000 increases the probability of home ownership by 2.0 percentage
points, and achieving a bachelors degree or more increases the probability by
1.5 percentage points. Understanding how these effects differ across the
quantiles is important from a policy standpoint. For instance, if
policymakers are looking to push more people into home ownership, they can
consider the various types of people (high utility - low utility), and focus
policy on the variables that have a greater impact on the subgroups.
Additionally, when downturns occur, there are clear difficulties
transitioning into or out of housing markets, which is clear from the results
of the interaction term. These results, along with those of the demographic
variables, shed light on findings that are unavailable or different than
those produced from modeling the mean (PBLD).

\section{Conclusion} \label{sec:Conclusion}

This paper presents quantile regression methods for binary longitudinal data
that accommodate various forms of heterogeneity, and designs an estimation
algorithm to fit the model. The framework developed in this paper contributes
to literatures on quantile regression for discrete data, panel data models
for quantile regression, and discrete panel data models. A simulation study
is performed, which demonstrates the computational efficiency of the
estimation algorithm and blocking approach.

The model is first applied to examine female labor force participation.
Although this is a heavily studied topic, the panel quantile approach offers
a new perspective to understand the impact of the covariates, while
controlling for heterogeneity and state dependence. The results show that
particular attention needs to be paid to women with newborns and children
ages 3-5 as the impacts of these variables on female labor force
participation are large and dispersed across the quantiles. The model is also
applied to investigate the determinants of home ownership before, during, and
after the Great Recession. The state dependence effect in home ownership is
strong (even when controlling for multivariate heterogeneity), however, after
the recession the effect differs nontrivially from mean regression. Other
results, including race, number of children, gender, health insurance, and
location, also offer unique findings across the quantiles, which are
unavailable in other modeling settings. The approach provided in this paper
leads to a richer view of how the covariates influence the outcome variables,
which better informs policy on female labor force participation and home
ownership.

\clearpage \pagebreak
\pdfbookmark[1]{References}{unnumbered}       
\section*{References}

\bibliography{EAQRBLDbib}
\bibliographystyle{jasa}

\clearpage
\newpage \appendix \renewcommand\thesection{Appendix \Alph{section}}

\section{Non-blocked Sampling in QBLD Model}\label{app:A}

The algorithm below presents the sampler for non-blocked sampling in the QBLD
model.
\begin{table*}[tbh]
\begin{algorithm}[Non-blocked sampling]
\label{alg:Non-block} \rule{\textwidth}{0.5pt} \small{
\begin{enumerate}[itemsep=-2ex]
\item Let $\Psi_{i} = D_{\tau \sqrt{w_{i}}}^{2}$. Sample $\beta|\alpha,\varphi^{2},
z,w$ $\sim$  $N(\tilde{\beta}, \tilde{B})$, where,
     \begin{equation*}
     \tilde{B}^{-1} = \bigg(\sum_{i=1}^{n}
     X'_{i} \Psi_{i}^{-1}   X_{i}
     + B_{0}^{-1} \bigg) \enskip
     \mathrm{and} \enskip
     \tilde{\beta} = \tilde{B}\left(\sum_{i=1}^{n} X'_{i} \Psi_{i}^{-1}
     (z_{i} - S_{i} \alpha_{i} - \theta w_{i}) + B_{0}^{-1}\beta_{0} \right).
     \end{equation*}
\item Sample $\alpha_{i}| \beta,\varphi^{2},z,w $ $\sim$  $N(\tilde{a}, \tilde{A})$
     for $i=1,\cdots,n$, where,
     \begin{equation*}
     \tilde{A}^{-1} = \left(S'_{i} \, D^{-2}_{\tau \sqrt{w_{i}}} \, S_{i}
     + \frac{1}{\varphi^{2}}  I_{l}\right)
     \quad \mathrm{and} \quad
     \tilde{a} = \tilde{A} \left(S'_{i} D^{-2}_{\tau \sqrt{w_{i}}} \,
     \big(z_{i} - X_{i} \beta - \theta w_{i} \big)   \right).
     \end{equation*}
\item    Sample $w_{it}|\beta, \alpha_{i}, z_{it}$ $\sim$  $GIG \, (0.5, \tilde{\lambda}_{it},
         \tilde{\eta})$ for $i=1,\cdots,n$ and $t=1,\cdots,T_{i}$, where,
         \begin{equation*}
         \tilde{\lambda}_{it} = \bigg( \frac{ z_{it} - x'_{it}\beta - s'_{it}\alpha_{i}}{\tau}
         \bigg)^{2} \quad \mathrm{and} \quad \tilde{\eta} = \bigg(
         \frac{\theta^{2}}{\tau^{2}} + 2 \bigg).
         \end{equation*}
\item    Sample $\varphi^{2}|\alpha \sim IG(\tilde{c}_{1}/2, \tilde{d}_{1}/2)$, where
         $\tilde{c}_{1} = \Big( nl + c_{1} \Big)$ and
         $\tilde{d}_{1} = \Big( \sum_{i} \alpha_{i}'\alpha_{i} + d_{1} \Big)$. \\
\item    Sample the latent variable $z|y,\beta,\alpha,w$ for
         all values of $i=1,\cdots,n$ and $t=1, \cdots, T_{i}$ from an univariate
         truncated normal (TN) distribution as follows,
         \begin{eqnarray*}
         z_{it}|y, \beta,w & \sim & \left\{
         \begin{array}{ll}
         TN_{(-\infty,0]}\bigg(x'_{it}\beta + s'_{it}\alpha_{i} + \theta w_{it},
         \tau^{2} w_{it} \bigg)
         & \textrm{if} \;\;  y_{it} = 0,\\ [0.9em]
         TN_{(0,\infty)}\bigg(x'_{it}\beta + s'_{it} \alpha_{i} + \theta w_{it},
         \tau^{2} w_{it} \bigg)
         & \textrm{if} \;\; y_{it} = 1.
         \end {array}
         \right.
         \end{eqnarray*}
\end{enumerate}}
\rule{\textwidth}{0.5pt}
\end{algorithm}
\end{table*}

\section{The Conditional Densities for Blocked Sampling in QBLD Model}
\label{app:B}

This appendix presents a derivation of the conditional posterior densities
for blocked sampling in the QBLD model. Specifically, the parameters $\beta$
and latent variable $z_i$ are sampled marginally of the random effects
parameter $\alpha_i$, from an updated multivariate normal and a truncated
multivariate normal distribution, respectively. The parameter $\alpha_i$ is
sampled conditional on ($\beta,z_i$) from an updated multivariate normal
distribution. The latent weights $w$ are sampled element wise from a
generalized inverse Gaussian (GIG) distribution and the variance
$\varphi^{2}$ is sampled from an updated inverse-gamma distribution.

\textbf{(1)}. The mean and variance of the QBLD model, $z_{i} = X_{i} \beta +
S_{i} \alpha_{i} + \theta w_{i} + D_{\tau \sqrt{w_{i}}} \, u_{i}$ for
$i=1,\ldots,n$, (marginally of $\alpha_{i}$) can be shown to have the
following expressions,
\begin{eqnarray*}
E(z_{i}) & = & X_{i} \beta + \theta w_{i}, \\
V(z_{i}) & = & \varphi^{2} S_{i} S'_{i} + D^{2}_{\tau \sqrt{w_{i}}} = \Omega_{i}.
\end{eqnarray*}
First, we derive the conditional posterior of $\beta$ and $z_i$, marginally
of $\alpha_i$, but conditional on other variables in the model.

\textbf{1(a)}. Starting with $\beta$, the conditional posterior density
$\pi(\beta|z,w,\varphi^{2})$ can be derived as,
\begin{align*}
\pi(\beta|z,w,\varphi^{2}) & \propto  \bigg\{ \prod_{i=1}^{n}
f(z_{i}|\beta,w_{i},\varphi^{2}) \bigg\} \; \pi(\beta) \\
& \propto \exp\bigg[ -\frac{1}{2} \bigg\{ \sum_{i=1}^{n} (z_{i} - X_{i}\beta -
\theta w_{i})'\Omega_{i}^{-1}(z_{i} - X_{i}\beta - \theta w_{i}) \\
& \quad + (\beta - \beta_{0})'B_{0}^{-1}(\beta - \beta_{0})\bigg\}   \bigg]\\
& \propto \exp\bigg[ -\frac{1}{2} \bigg\{ \beta'\bigg(\sum_{i=1}^{n} X'_{i}
\Omega_{i}^{-1} X_{i} + B_{0}^{-1} \bigg) \beta   -
\beta' \bigg( \sum_{i=1}^{n} X'_{i} \Omega_{i}^{-1} (z_{i} - \theta w_{i}) +
B_{0}^{-1} \beta_{0} \bigg) \\
& \quad -  \bigg( \sum_{i=1}^{n} (z_{i} - \theta w_{i})'\Omega_{i}^{-1} X_{i}
+ \beta'_{0} B_{0}^{-1}  \bigg)  \bigg\}   \bigg] \\
& \propto \exp \bigg[ -\frac{1}{2} \bigg\{ \beta' \tilde{B}^{-1} \beta -
\beta' \tilde{B}^{-1} \tilde{\beta} - \tilde{\beta}' \tilde{B}^{-1} \beta
\bigg\}  \bigg],
\end{align*}
where the third line only keeps terms involving $\beta$ and the fourth line
introduces the terms $\tilde{\beta}$ and $\tilde{B}$, which are defined as,
\begin{equation*}
\tilde{B}^{-1} = \bigg(\sum_{i=1}^{n} X'_{i} \Omega_{i}^{-1} X_{i} + B_{0}^{-1} \bigg)
\quad \mathrm{and} \quad \tilde{\beta}  = \tilde{B}
\bigg(X'_{i}\Omega_{i}^{-1}(z_{i}-\theta w_{i}) + B_{0}^{-1} \beta_{0}\bigg).
\end{equation*}
Adding and subtracting $\tilde{\beta}'\tilde{B}^{-1}\tilde{\beta}$ and
absorbing the term $\exp[-\frac{1}{2}
\{-\tilde{\beta}'\tilde{B}^{-1}\tilde{\beta}\}]$ into the proportionality
constant, the square can be completed as follows,
\begin{align*}
\pi(\beta|z,w,\varphi^{2}) & \propto \exp\bigg[ -\frac{1}{2}
(\beta - \tilde{\beta})' \tilde{B}^{-1} (\beta - \tilde{\beta})  \bigg].
\end{align*}
The above expression is recognized as the kernel of a Gaussian or normal
distribution and hence $\beta|z,w,\varphi^{2} \sim
N(\tilde{\beta},\tilde{B})$.

\textbf{1(b)}. The conditional posterior density of the latent variable $z$
marginally of $\alpha$ can be obtained from the joint posterior density
\eqref{eq:jointPosterior} as,
\begin{align*}
\pi(z|\beta,w,\varphi^{2},y) & \propto  \prod_{i=1}^{n}
\bigg\{ \pi(z_{i}|\beta,w_{i},\varphi^{2},y_{i})  \bigg\} \\
& \propto \prod_{i=1}^{n} \bigg\{ \prod_{t=1}^{T_{i}}
\Big[I(z_{it}>0) I(y_{it}=1) + I(z_{it} \leq 0) I(y_{it}=0) \Big]  \\
& \quad \times
\exp\bigg[-\frac{1}{2} (z_{i}-X_{i}\beta - \theta w_{i})' \Omega_{i}^{-1}
(z_{i}-X_{i}\beta - \theta w_{i})   \bigg]\bigg\}.
\end{align*}
The expression inside the curly braces corresponds to a truncated
multivariate normal distribution, so $z_{i}|y_{i},\beta,w_{i},\varphi^{2}
\sim TMVN_{B_{i}}(X_{i}\beta + \theta w_{i}, \Omega_{i})$ for all
$i=1,\cdots,n$. Here, $B_{i}$ is the truncation region such that $B_{i} =
(B_{i1} \times B_{i2} \times \ldots \times B_{iT_{i}})$, where $B_{it}$ is
the interval $(0,\infty)$ if $y_{it}=1$ and the interval $(-\infty,0]$ if
$y_{it}=0$ for $t=1,\ldots,T_{i}$. Sampling directly from a TMVN is not
possible, hence we resort to the method proposed in \citet{Geweke-1991},
which utilizes Gibbs sampling to make draws from a TMVN.

Let $z_{i}^{j}$ denote the values of $z_{i}$ at the $j$-th pass of the MCMC
iteration. Then sampling is done from a series of conditional posterior
distribution as follows:
\begin{equation*}
z_{it}^{j}|z_{i1}^{j},\cdots,z_{i(t-1)}^{j},z_{i(t+1)}^{j},\cdots,z_{iT_{i}}^{j}
\sim  TN_{B_{it}}(\mu_{t|-t}, \Sigma_{t|-t}), \hspace{0.5in} \mathrm{for}
\; t=1,\cdots,T_{i},
\end{equation*}
where $TN$ denotes a truncated normal distribution. The terms $\mu_{t|-t}$
and $\Sigma_{t|-t}$ are the conditional mean and variance, respectively, and
are defined as,
\begin{eqnarray*}
\mu_{t|-t} & = & x'_{it}\beta + \theta w_{it} + \Sigma_{t,-t} \Sigma_{-t,-t}^{-1}
\big( z_{i,-t}^{j} - (X_{i}\beta + \theta w_{i})_{-t} \big), \\
\Sigma_{t|-t} & = & \Sigma_{t,t} - \Sigma_{t,-t} \Sigma_{-t,-t}^{-1}\Sigma_{-t,t},
\end{eqnarray*}
where $z_{i,-t}^{j}=(z_{i1}^{j}, \cdots,z_{i(t-1)}^{j},z_{i(t+1)}^{j-1},
\cdots,z_{iT_{i}}^{j-1})$, $(X_{i}\beta + \theta w_{i})_{-t}$ is a column
vector with $t$-th element removed, $\Sigma_{t,t}$ denotes the $(t,t)$-th
element of $\Omega_{i}$, $\Sigma_{t,-t}$ denotes the $t$-th row of
$\Omega_{i}$ with element in the $t$-th column removed and $\Sigma_{-t,-t}$
is the $\Omega_{i}$ matrix with $t$-th row and $t$-th column removed.

\textbf{(2)}. The conditional posterior density of the random effects
parameters $\alpha_{i}$ for $i=1,\ldots,n$ is derived from the joint
posterior density \eqref{eq:jointPosterior} as follows,
\begin{align*}
\pi(\alpha_{i}|z_{i},\beta,w_{i},\varphi^{2}) & \propto
f(z_{i}|\beta,\alpha_{i},w_{i}) \; \pi(\alpha_{i}|\varphi^{2}) \\
& \propto \exp\bigg[-\frac{1}{2} \bigg\{ (z_{i} - X_{i}\beta - S_{i}\alpha_{i}-\theta w_{i})'
D^{-2}_{\tau \sqrt{w_{i}}}  (z_{i} - X_{i}\beta - S_{i}\alpha_{i}-\theta w_{i}) \\
& \quad  + \frac{ \alpha'_{i} \alpha_{i}}{\varphi^{2}} \bigg\}  \bigg] \\
& \propto \exp\bigg[ -\frac{1}{2} \bigg\{ \alpha'_{i}
\Big(S'_{i} D^{-2}_{\tau \sqrt{w_{i}}} S_{i} +  \varphi^{-2} I_{l} \Big) \alpha_{i}
- \alpha'_{i} \Big( S'_{i} D^{-2}_{\tau\sqrt{w_{i}}} \big(z_{i} - X_{i}\beta
-\theta w_{i} \big) \Big) \\
& \quad - \Big( \big(z_{i} - X_{i}\beta
-\theta w_{i} \big)' D^{-2}_{\tau\sqrt{w_{i}}} S_{i}\Big) \alpha_{i}
\bigg\}  \bigg] \\
& \propto \exp\bigg[-\frac{1}{2} (\alpha_{i} - \tilde{a})'\tilde{A}^{-1}
(\alpha_{i} - \tilde{a})  \bigg],
\end{align*}
where the third line omits all terms not involving $\alpha_{i}$ and the
fourth line introduces the terms,
 \begin{equation*}
 \tilde{A}^{-1} = \left(S'_{i} \, D^{-2}_{\tau \sqrt{w_{i}}} \, S_{i}
 + \frac{1}{\varphi^{2}}  I_{l}\right)
 \quad \mathrm{and} \quad
 \tilde{a} = \tilde{A} \left(S'_{i} D^{-2}_{\tau \sqrt{w_{i}}} \,
 \big(z_{i} - X_{i} \beta - \theta w_{i} \big)   \right),
 \end{equation*}
as the posterior precision and posterior mean, respectively, and completes
the square. The result is a kernel of a normal distribution, hence,
$\alpha_{i}|z_{i},\beta,w_{i},\varphi^{2} \sim N(\tilde{a}, \tilde{A})$ for
$i=1,\ldots,n$.

\textbf{(3)}. The conditional posterior density of $w$ is obtained from the
joint posterior density \eqref{eq:jointPosterior} by collecting terms
involving $w$. Each term in $w$ is updated element-wise as follows:
\begin{align*}
\pi(w_{it}|z_{it},\beta,\alpha_{i}) & \propto
\big(2\pi \tau^{2} w_{it} \big)^{-1/2} \exp\bigg[ -\frac{1}{2 \tau^{2} w_{it}}
\big( z_{it} - x'_{it}\beta - s'_{it}\alpha_{i}-\theta w_{it}  \big)^{2}
- w_{it}  \bigg] \\
& \propto w_{it}^{-1/2} \exp\bigg[ -\frac{1}{2} \bigg\{
\bigg( \frac{ z_{it} - x'_{it}\beta - s'_{it}
 \alpha_{i}}{\tau} \bigg)^{2} w_{it}^{-1}
+ \bigg( \frac{\theta^{2}}{\tau^{2}} + 2  \bigg) w_{it} \bigg\}  \bigg] \\
& \propto w_{it}^{-1/2} \exp \bigg[ -\frac{1}{2} \bigg\{ \tilde{\lambda}_{it}
w_{it}^{-1} + \tilde{\eta} w_{it}  \bigg\}  \bigg],
\end{align*}
where the second line omits all terms not involving $w_{it}$ and the third
line introduces the terms defined below,
 \begin{equation*}
 \tilde{\lambda}_{it} = \bigg( \frac{ z_{it} - x'_{it}\beta - s'_{it}
 \alpha_{i}}{\tau} \bigg)^{2} \quad \mathrm{and} \quad \tilde{\eta}
 = \bigg(\frac{\theta^{2}}{\tau^{2}} + 2 \bigg).
 \end{equation*}
The expression in the third line is recognized as the kernel of a generalized
inverse Gaussian (GIG) distribution. Hence, we have
$w_{it}|z_{it},\beta,\alpha_{i} \sim GIG(0.5,\tilde{\lambda}_{it},
\tilde{\eta})$ for $t=1,\ldots,T_{i}$ and $i=1,\ldots,n$.

\textbf{(4)}. The conditional posterior density of $\varphi^{2}$ is obtained
from the joint posterior density \eqref{eq:jointPosterior} by collecting
terms involving $\varphi^{2}$ conditional on the remaining model parameters.
This is done below.
\begin{align*}
\pi(\varphi^{2}|\alpha) & \propto
(2\pi)^{-nl/2} \big(\varphi^{2}\big)^{-nl/2} \exp \bigg[ -\frac{1}{2\varphi^{2}}
\sum_{i=1}^{n} \alpha'_{i} \alpha_{i} \bigg] \big(\varphi^{2}\big)^{-(c_{1}/2 + 1)}
\exp\bigg[-\frac{d_{1}}{2 \varphi^{2}} \bigg] \\
& \propto \big(\varphi^{2}\big)^{-(nl/2 + c_{1}/2 + 1)}
\exp\bigg[-\frac{1}{2\varphi^{2}} \bigg\{ \sum_{i=1}^{n} \alpha'_{i}\alpha_{i}
+ d_{1} \bigg\}  \bigg] \\
& \propto \big(\varphi^{2}\big)^{(\tilde{c}_{1}/2 + 1)} \exp\bigg[
-\frac{1}{2 \varphi^{2}} \; \tilde{d}_{1}\bigg],
\end{align*}
where $\tilde{c}_{1} = nl + c_{1}$ and $\tilde{d}_{1} = \big(\sum_{i=1}^{n}
\alpha'_{i}\alpha_{i} + d_{1} \big)$. The expression in the last line is
recognized as the kernel of an inverse gamma (IG) distribution and
consequently, we have $\varphi^{2}|\alpha \sim IG(\tilde{c}_{1}/2,
\tilde{d}_{1}/2)$.

\end{document}